\documentclass{sig-alternate}
\usepackage{amsmath}
\usepackage{times}
\usepackage{color}
\usepackage{url}
\usepackage{graphicx}
\usepackage{pifont}
\usepackage{booktabs}
\usepackage{subfigure}
\usepackage{xspace}
\usepackage{epstopdf}
\graphicspath{{FIG/}}
\usepackage{array}

\newfont{\mycrnotice}{ptmr8t at 7pt}
\newfont{\myconfname}{ptmri8t at 7pt}
\let\crnotice\mycrnotice%
\let\confname\myconfname%

\permission{Permission to make digital or hard copies of all or part of this work for personal or classroom use is granted without fee provided that copies are not made or distributed for profit or commercial advantage and that copies bear this notice and the full citation on the first page. Copyrights for components of this work owned by others than ACM must be honored. Abstracting with credit is permitted. To copy otherwise, or republish, to post on servers or to redistribute to lists, requires prior specific permission and/or a fee. Request permissions from permissions@acm.org.}
\conferenceinfo{WSDM'14,}{February 24--28, 2014, New York, New York, USA.}
\copyrightetc{Copyright 2014 ACM \the\acmcopyr}
\crdata{978-1-4503-2351-2/14/02\ ...\$15.00.\\
http://dx.doi.org/10.1145/2556195.2556243}

\begin{document}

\newtheorem{theorem}{Theorem}
\newtheorem{corollary}{Corollary}
\newtheorem{fassumption}{Fundamental Assumption}
\newtheorem{lemma}{Lemma}
\newtheorem{definition}{Definition}
\newcommand{\cmark}{\ding{51}}%
\newcommand{\xmark}{\ding{55}}%
\newcommand{\hide}[1]{}
\newcommand{\note}[1]{{\textsf{\textcolor{red}{\ \ #1 \ }}}}
\newcommand{\xhdr}[1]{\vspace{1mm}\noindent{{\bf #1.}}}
\newcommand{\eg}{\emph{e.g.}}
\newcommand{\ie}{\emph{i.e.}}

\newcommand{\model}{CoDA\xspace}
\newcommand{\fullmodel}{Communities through Directed Affiliations\xspace}
\newcommand{\fullmodelCAPS}{COMMUNITIES THROUGH DIRECTED AFFILIATIONS}
\newcommand{\rev}[1]{\textbf{*** #1}}
\newcommand{\jure}[1]{\textbf{\color{red}{[[ #1 ]]}}}

\newcommand{\beq}{\begin{equation}}
\newcommand{\eeq}{\end{equation}}
\newcommand{\argmax}{\operatornamewithlimits{argmax}}
\newcommand{\argmin}{\operatornamewithlimits{argmin}}
\newenvironment{squeezelist}{
    \begin{list}{\tiny{$\bullet$}}%
    {\setlength{\itemsep}{1ex} \setlength{\topsep}{1ex}
    \setlength{\parsep}{0pt} \setlength{\itemindent}{0pt}
    \setlength{\leftmargin}{1.5em}
    \setlength{\partopsep}{1pt}}}%
    {\end{list}}


\title{
Detecting Cohesive and 2-mode Communities in\\ Directed and Undirected Networks
}
\author{
Jaewon Yang, Julian McAuley, Jure Leskovec\\
Stanford University\\
\{jayang, jmcauley, jure\}@cs.stanford.edu
}

\maketitle

\begin{abstract}
Networks are a general language for representing relational information among objects. An effective way to model, reason about, and summarize networks, is to discover sets of nodes with common connectivity patterns. Such sets are commonly referred to as {\em network communities}. Research on network community detection has predominantly focused on identifying communities of densely connected nodes in undirected networks.

In this paper we develop a novel overlapping community detection method that scales to networks of millions of nodes and edges and
advances research along two dimensions: the connectivity structure of communities, and the use of edge directedness for community detection.
First, we extend traditional definitions of network communities by building on the observation that nodes can be densely interlinked in two different ways: In {\em cohesive} communities nodes link to each other, while in {\em 2-mode} communities nodes link in a bipartite fashion,
where links predominate \emph{between} the two partitions rather than inside them.
Our method successfully detects both 2-mode as well as cohesive communities, that may also overlap or be hierarchically nested. Second, while most existing community detection methods treat directed edges as though they were undirected, our method accounts for edge directions and is able to identify novel and meaningful community structures in both directed and undirected networks, using data from social, biological, and ecological domains.

\end{abstract}

\noindent
{\bf Categories and Subject Descriptors:} H.2.8 {\bf [Database Management]}: {Database Applications -- {\em Data mining}}

\noindent
{\bf General Terms:} Algorithms, theory, experimentation.

\vspace{-.2cm}

\section{Introduction}
\label{sec:intro}



Networks are a powerful way to model relational information among objects from social, natural, and technological domains. Networks can be studied at various levels of resolution ranging from whole networks to individual nodes.
Arguably the most useful level of resolution is
at the level of groups of nodes. Studying groups of nodes allows us to identify and analyze modules or components of networks. For example, understanding the organization of networks at the level of groups helps us to discover functional roles of proteins in protein-protein interaction networks~\cite{Pinkert10ProteinModules}, political factions in a network of bloggers~\cite{adamicBlogs}, social circles in online social networks~\cite{nips2012}, or even topics in word association networks~\cite{Ahn10LinkCommunitiesNature}.

One way to understand networks at the level of groups is to identify sets of nodes with similar connectivity patterns. Traditional methods aim to find network \emph{communities}, which are defined as groups of nodes with many connections among the group's members, but few to the rest of the network~\cite{Ahn10LinkCommunitiesNature,demon,fortunato09community, palla05_OveralpNature}. However, dense communities are but one kind of group structure in networks, and there may be other structures that help us to understand networks better.
For example, consider a Twitter follower network and the ``community'' of candidates in the 2012 U.S. presidential election. This community is not densely interlinked, in the sense that the candidates 
do not follow each other; thus we would not be able to find this community if we were to use traditional methods that search for densely connected sets of nodes. However, such communities can be identified because they form around nodes whose edges have similar endpoints. Continuing our example, presidential candidates form a community in Twitter not because they follow each other but because a common set of ``fans'' follows them.

\begin{figure}[t]
	\hspace{5mm}\includegraphics[width=0.75\linewidth]{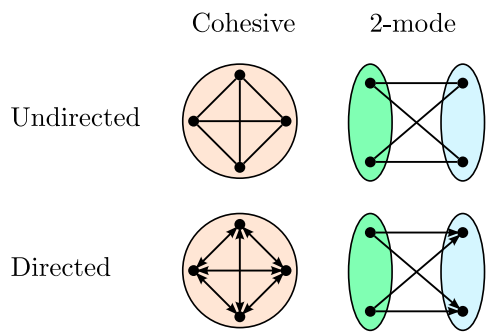}
	\vspace{-1mm}
  \caption{Two types of networks (directed and undirected) and two types of communities (cohesive and 2-mode). While research has predominantly focused on undirected-cohesive communities (top left), we develop a method that can detect cohesive as well as 2-mode communities in both directed and undirected networks.}
	\label{fig:example}
	\vspace{-3mm}
\end{figure}

Thus communities can be characterized by the connectivity structure between the members {\em and also} by the connectivity structure of the members to the rest of the network. We refer to these communities
as {\em 2-mode communities}. For example, in case of ``fans'' linking to ``celebrities'' members of a community may be linked to the same set of endpoints, even if they do not link to each other.
%
Similar examples also exist beyond social networks; for example, in protein-protein interaction networks, some protein complexes act as bridges or regulators, \ie, they do not interact among themselves but regulate/interact with the same set of proteins~\cite{Pinkert10ProteinModules}.

Another common assumption made by many present community detection methods is that networks are \emph{undirected}~\cite{palla05_OveralpNature,jaewon12agmfit}. This implies that relationships between connected nodes are symmetric or reciprocal. However, in directed networks relationships are asymmetric, as with our previous example about ``fans'' who follow ``celebrities''. Even though methods can often be adapted to handle directed networks, this is often done in an ad-hoc fashion (\eg, by treating directed edges as though they were undirected) and can lead to unexpected or undesirable results~\cite{fortunato09community,Newman08DirectedCommunity,rosvall08infomap}.
Moreover, by ignoring edge directedness important information may be lost, especially if relationships are predominantly non-reciprocal as in predator-prey networks~\cite{Newman08DirectedCommunity} or in social networks like Twitter. 




\xhdr{Present work: Detecting cohesive and 2-mode communities in directed and undirected networks}
Here we consider new notions of community linking structure that go beyond thinking of communities as internally well-connected sets of nodes.
%
Our work stems from social network literature on structural equivalence~\cite{burt78}, where it has been noted that social homogeneity (\ie, social communities) arises not only between nodes that link to each other (\ie, {\em internal group connectivity}), but also between nodes that link to the rest of the network in a coordinated way (\ie, {\em external group connectivity}).
In particular, we consider different notions of ``communities'' that
are depicted in Figure \ref{fig:example}. We differentiate between {\em cohesive} communities (Fig.~\ref{fig:example}, Cohesive) and {\em 2-mode} communities (Fig.~\ref{fig:example}, 2-mode) where nodes link in a bipartite fashion with links predominantly appearing between partitions rather than inside them.

While existing community detection methods typically focus on Undirected-Cohesive or Directed-Cohesive communities~\cite{demon,fortunato09community,Newman08DirectedCommunity,palla05_OveralpNature,jaewon12agmfit},
the focus of our paper is on developing methods that can detect communities of all four different types depicted in Figure~\ref{fig:example}. By modeling each of these definitions in concert, we are able to capture the complex structure present in networks.

\xhdr{Present work: \fullmodel}
We present {\em \model (\fullmodel)}, a meth\-od for overlapping community detection that scales to networks with millions of nodes and tens of millions of edges. \model exhibits the following three properties: (1) It naturally detects both cohesively connected as well as 2-mode communities. (2) \model allows cohesive and 2-mode communities to overlap or be hierarchically nested. (3) \model naturally allows for community detection in directed as well as undirected networks.

We develop our community detection method by first presenting a generative model of networks where edges arise from affiliations of nodes to cohesive and 2-mode communities. Then we fit the model to a given network and thus discover communities.

Our model starts with a bipartite affiliation graph \cite{Lattanzi09AffiliationNetworks,jaewon12agmfit,jaewon13agmfast,zheleva09affiliation},
where nodes of the underlying network represent one `layer' of the bipartite graph and communities represent the other. Edges between network-nodes and community-nodes in the affiliation graph represent memberships of nodes to communities. However, our approach has a simple but critical innovation: while memberships of nodes to communities have previously been modeled as undirected, we model the memberships as {\em directed}.

Though simple on the surface, this modification leads to substantial changes in the modeling capability of affiliation network models.
In particular, a directed affiliation between a node and a community models whether the node \emph{sends} or \emph{receives} (or both) links to other members of the community. Directed affiliations allow us to simultaneously model cohesive as well as 2-mode communities. In cohesive communities node affiliations are bidirectional (a node both sends \emph{and} receives links from other members); 2-mode communities are modeled with unidirectional memberships where some members mostly send/create links (\ie, fans) while others mostly receive them (\ie, celebrities).


Having defined the node-community affiliation model we then develop a method to fit the model to a given network. Our model fitting procedure builds on that of the BigCLAM community detection method~\cite{jaewon13agmfast}.
Although we solve a more complex problem than BigCLAM (\ie, we find both 2-mode as well as cohesive communities), we employ similar approximation techniques.
Until recently, methods for overlapping community detection could only process networks with up to around 10,000 nodes~\cite{mmsbScale12}.
In contrast, \model can easily handle networks that are two orders of magnitude larger: millions of nodes, tens of millions of edges. Moreover, \model can be easily parallelized which further increases the scalability.

\xhdr{Present work: Experimental results}
We evaluate \model on a number of networks from various domains. We consider social, biological, communication, and ecological networks. We test \model on networks with explicitly labeled ground-truth communities~\cite{nips2012,jaewon11comscore} as well as on networks where communities can be manually examined.

Experiments demonstrate that \model's ability to detect 2-mode as well as cohesive communities leads to improved performance over the existing state-of-the-art.
For example, when detecting social circles in the Google+ online social network, \model gives a relative improvement in accuracy of 36\% over Link clustering~\cite{Ahn10LinkCommunitiesNature} (28\% over MMSB~\cite{airoldi07blockmodel}, 25\% over clique percolation~\cite{palla05_OveralpNature} and 21\% over DEMON~\cite{demon}).


More importantly, \model facilitates novel discoveries about the community structure of networks. For example, we find that 2-mode communities in foodwebs of predatory relations between organisms correspond to groups of predators who rely on similar groups of prey.
Interestingly we find that in scientific paper citation networks, protein-protein interaction networks, as well as web graphs, the majority of detected communities are 2-mode. However, in social networks where edges signify reciprocal friendships, cohesive communities are more frequent. In Twitter or Google+, where relationships are asymmetric, 2-mode communities represent a significant portion of the network (20\% in Twitter and 30\% in Google+).

\xhdr{Further related work}
%
While there exist a number of different definitions of network communities~\cite{fortunato09community}, traditionally, communities have been thought of as densely connected sets of nodes~\cite{Ahn10LinkCommunitiesNature,danon05community,palla05_OveralpNature,RCCLP04_PNAS}. In contrast, the notion of structural equivalence suggests that nodes with similar connectivity patterns may be considered a community even if they do not link to each other~\cite{burt78,Henderson12RolX,KRRT99_trawling}. Our work here builds on both notions of network communities and attempts to resolve them by using a single, unified model.

Detecting communities of densely connected sets of nodes is an extensively researched area~\cite{fortunato09community,Malliaros13Survey,papadopoulos11community,Xie13SurveyOverlapping} with a plethora of different algorithms and heuristics. For example, separate methods have been proposed for detecting communities in undirected networks that are disjoint \cite{andersen06seed,dhillon07graclus,karypis98metis,rosvall07_informationPNAS,Satuluri08Markov}, overlapping~\cite{airoldi07blockmodel,Ball11overlappingcommunities,demon,palla05_OveralpNature,jaewon12agmfit}, or hierarchically nested~\cite{Ahn10LinkCommunitiesNature,jaewon11comscore}.
%
%
On the other hand, detection of 2-mode communities has been much less researched. An exception here is {\em Trawling}~\cite{KRRT99_trawling}, which is a method for extracting 2-mode communities in large directed networks. The critical difference with our work here is that Trawling only identifies {\em complete} bipartite subgraphs of a given directed network. In contrast, our method is able to identify cohesive as well as bipartite communities in directed as well as undirected networks.

Conceptually \model is related to existing work on \emph{block models}, which are in principle capable of detecting cohesive as well as 2-mode communities~\cite{airoldi07blockmodel,mmsbScale12,Holland83BlockModel}. Our work differs from such approaches in terms of how communities overlap and are hierarchically nested. We also emphasize the scalability of \model compared to these approaches.

\model is an example of an affiliation network model~\cite{Lattanzi09AffiliationNetworks,jaewon12agmfit,jaewon13agmfast,zheleva09affiliation}. While existing affiliation network models can only model undirected cohesive communities, the crucial difference here is our ability to model \emph{directed} networks and \emph{2-mode} communities.

The rest of the paper is organized as follows.
Section~\ref{sec:proposed} defines the affiliation network model and Section~\ref{sec:fitting} discusses the model fitting procedure. We present experimental results in Sections~\ref{sec:experiments} and~\ref{sec:discussion}, and conclude in Section
\ref{sec:conclusion}.

\section{Directed Community\\ Affiliations}
\label{sec:proposed}

We start by presenting a stochastic generative model of networks in which the probability of an edge appearing between a pair of nodes depends on the community affiliations of thise nodes. We then develop an efficient model fitting procedure which allows for detecting community affiliations of nodes in a given network.

We describe our model in the context of directed networks and then show how it can straightforwardly be adapted to undirected networks. Our model builds upon BigCLAM, an affiliation model for overlapping network communities~\cite{jaewon13agmfast}. However, whereas BigCLAM focuses on finding only \emph{cohesive} communities in {\em undirected} networks, our work here aims to find 2-mode communities \emph{as well as} cohesive communities in both directed and undirected networks.



\xhdr{Directed Affiliation Network Model}
We begin with the intuition that a desirable model of communities in directed networks should exhibit two properties. First, communities should be modeled not only in terms of their internal connectivity, but also in terms how members connect to non-members. Second, the model should account for asymmetries, \ie, directedness, of edges between nodes. We later demonstrate that accounting for these two properties is important. Perhaps surprisingly, our method gives improved performance even when modeling communities in \emph{undirected} networks. This is due to the fact that when edge directions are not explicit, relationships in the network may still be (implicitly) asymmetric, and identifying such asymmetries leads to improved performance.

We proceed by formulating a simple conceptual model of networks that we refer to as a {\em Directed Affiliation Network Model}. Our work builds on a family of affiliation network models~\cite{breiger74groups}, however, existing affiliation models are typically designed to handle cohesive communities in undirected networks~\cite{Lattanzi09AffiliationNetworks,jaewon12agmfit,jaewon13agmfast,zheleva09affiliation}; here we extend such models in order to capture cohesive as well as 2-mode communities in directed as well as undirected networks.

To represent node community memberships, we consider a bipartite affiliation graph where the nodes of the network (bottom layer) connect to communities (top layer) to which they belong (Figure~\ref{fig:dagm.bipartite}). Edges of the underlying network (Figure~\ref{fig:dagm.network}) then arise due to shared community affiliations of nodes.

Consider for a moment an undirected network; when a node belongs to a community in such a network it typically means that the node has (undirected) edges to other members of the community. This type of community affiliation can be modeled using a bipartite graph of nodes and communities where undirected affiliations are formed between nodes and communities~\cite{Lattanzi09AffiliationNetworks,jaewon12agmfit,jaewon13agmfast,zheleva09affiliation}.

In directed networks, however, we need a richer notion of community affiliation (Figure~\ref{fig:dagm.bipartite}): a node may {\em create} edges \emph{to} other members of a community, and it also {\em receive} edges \emph{from} other members of the community, or both. Therefore, we assume that nodes in directed networks can have two ``types'' of community affiliation: ``Outgoing'' affiliations from nodes to communities mean that in the network the node {\em sends} edges to other members of the community. And, ``incoming'' affiliations from communities to nodes mean that nodes {\em receive} edges from other community members. We model this using {\em directed} memberships between nodes and communities: outgoing memberships and incoming memberships.

\begin{figure}[t]
\centering
  \subfigure[Node community affiliations]{\label{fig:dagm.bipartite}\parbox{43mm}{\centering\includegraphics[height=0.12\textwidth]{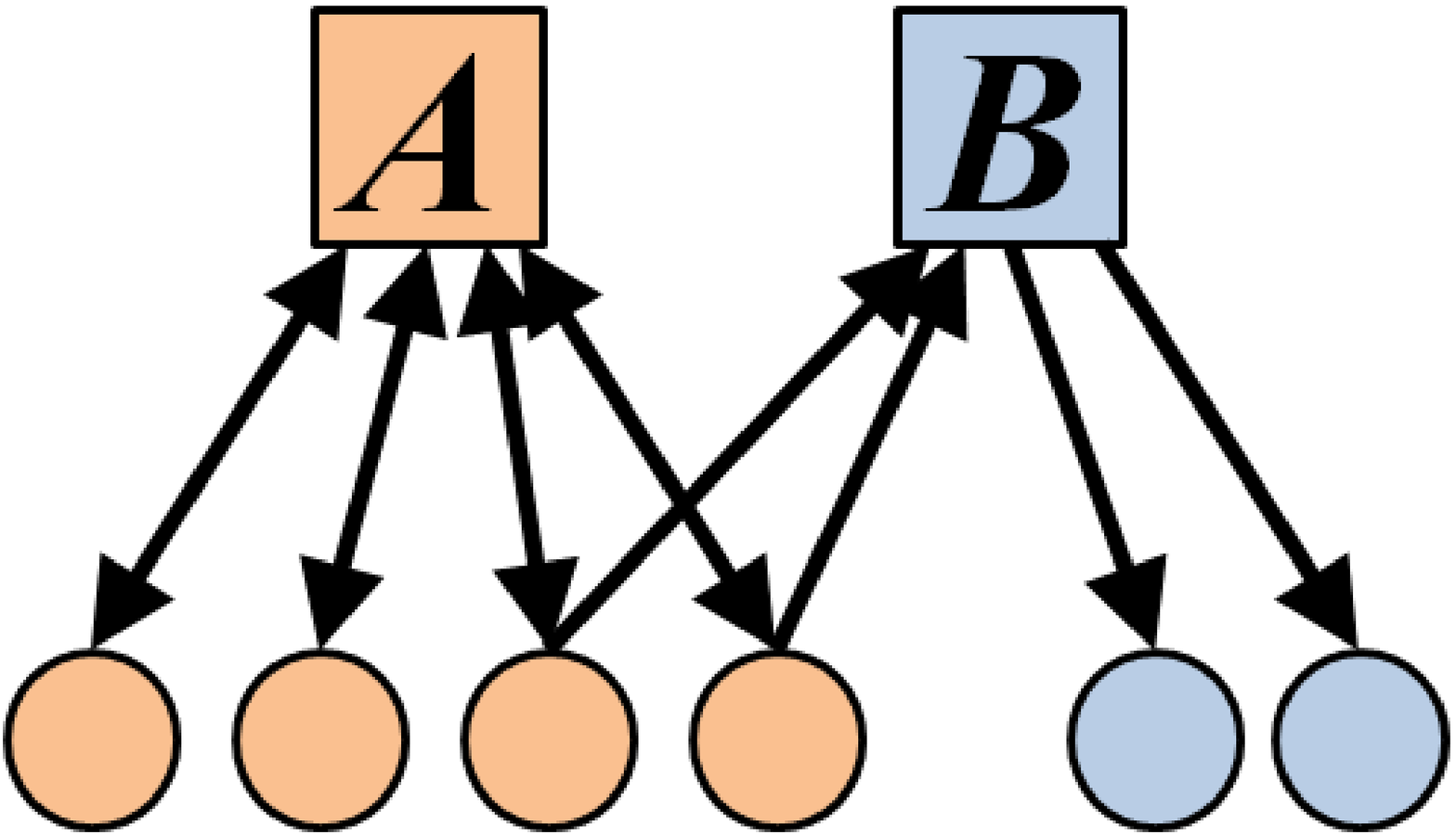}\vspace{3mm}}}
  \subfigure[Network $G$]{\label{fig:dagm.network}\parbox{40mm}{\centering\includegraphics[height=0.12\textwidth]{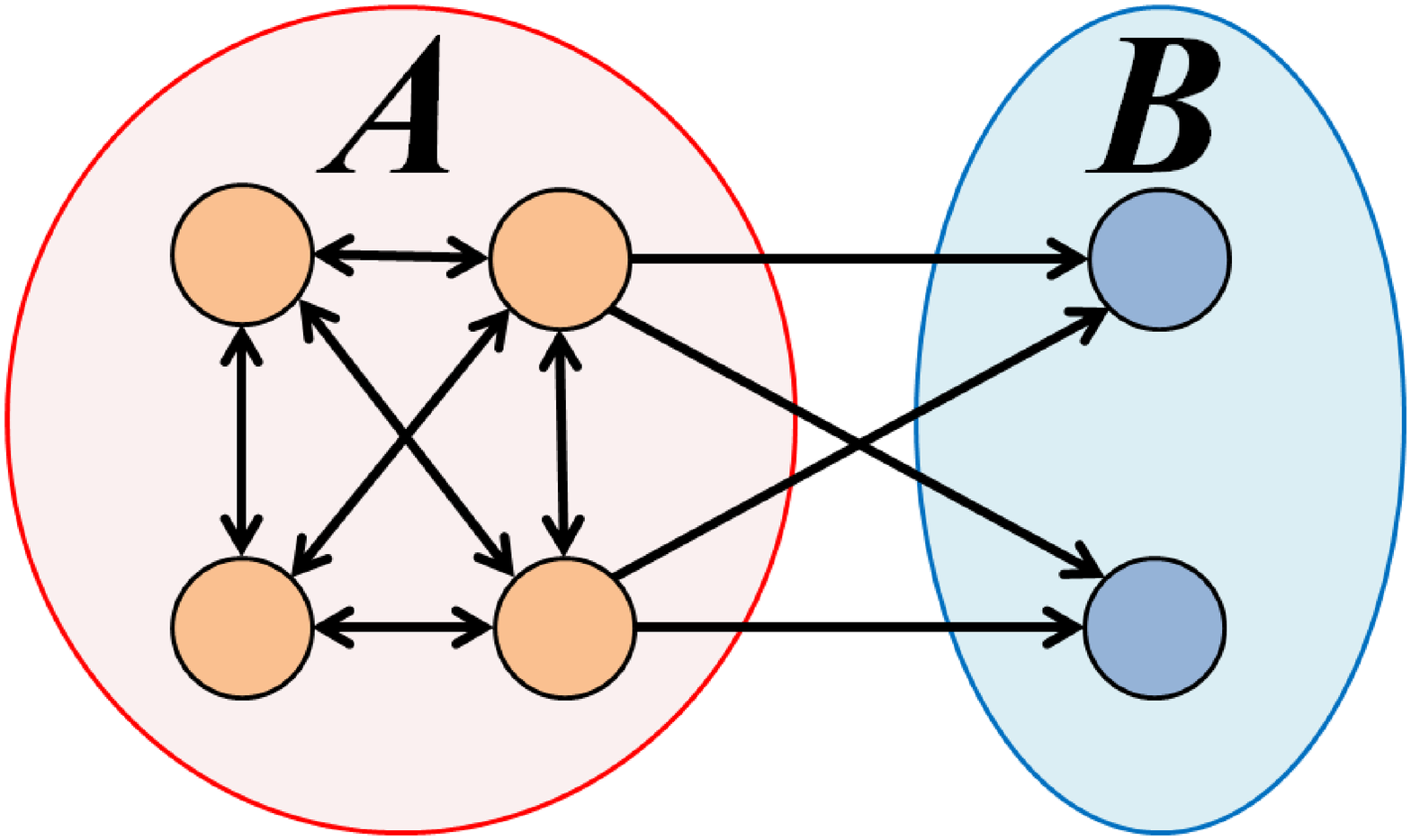}\vspace{3mm}}}
  \vspace{-2mm}
  \caption{
  (a) Directed node community affiliation graph. Squares: communities, Circles: nodes of network $G$. Affiliations from nodes to communities indicate that nodes \emph{create} edges to other members in those communities, while affiliations from communities to nodes indicate that nodes \emph{receive} edges from others. Community $A$ is cohesive, while $B$ is a 2-mode community.
  (b) Network $G$ corresponding to model in (a).
  }
  \vspace{-0mm}
  \label{fig:network.illustration}
\end{figure}


Formally, we denote a bipartite affiliation graph as $B(V,C,M)$, where $V$ is the set of nodes of the underlying network $G$, $C$ the set of communities, and $M$ a set of directed edges connecting nodes $V$ and communities $C$. An outgoing membership edge of node $u \in V$ to community $c \in C$ is denoted as $(u, c) \in M$ and, and an incoming membership is denoted as $(c, u) \in M$.

Now, given the affiliation graph $B(V,C,M)$, we need to specify a process that generates the edges $E$ of the underlying directed network $G(V,E)$. 
To this end we consider a simple parameterization where we assign a single parameter $p_c$ to every community $c \in C$. The parameter $p_c$ models the probability of a directed edge forming from a member node $u$ with an \emph{outgoing} membership to community $c$ to another member $v$ of $c$ with an \emph{incoming} membership.
In other words, we generate a directed edge between a pair of nodes with probability $p_c$ if they are connected in $B$ with a 2-step directed path via community $c$. Each community $c$ creates edges independently. However, if two nodes are connected by more than one community, duplicate edges are not included in the graph $G(V,E)$.

\begin{definition}[Directed Affiliation Network Model]
Let $B(V,C,M)$ be a directed bipartite graph where $V$ is a set of nodes, $C$ is a set of communities, and $M$ is a set of directed edges between $V$ and $C$.
Also, let $\{p_c\}$ be a set of probabilities for all $c \in C$. Given $B(V,C,M)$ and $\{p_c\}$, the model generates a directed graph $G(V,E)$ by creating a directed edge $(u,v)$ from node $u \in V$ to node $v \in V$ with probability $p(u,v)$:
\beq
p(u,v)= 1 - \prod_{k \in C_{uv}} (1 - p_k),
\label{eq:puv}
\eeq
where $C_{uv} \subset C$ is a set of communities through which $u$ has a 2-step directed path to $v$ ($C_{uv} = \{c | (u,c), (c, v) \in M\}$). If $C_{uv}=\varnothing$ then we set $p(u,v)=1/|V|$.
\end{definition}

\begin{figure}[t]
\begin{center}
 \includegraphics[width=0.77\linewidth]{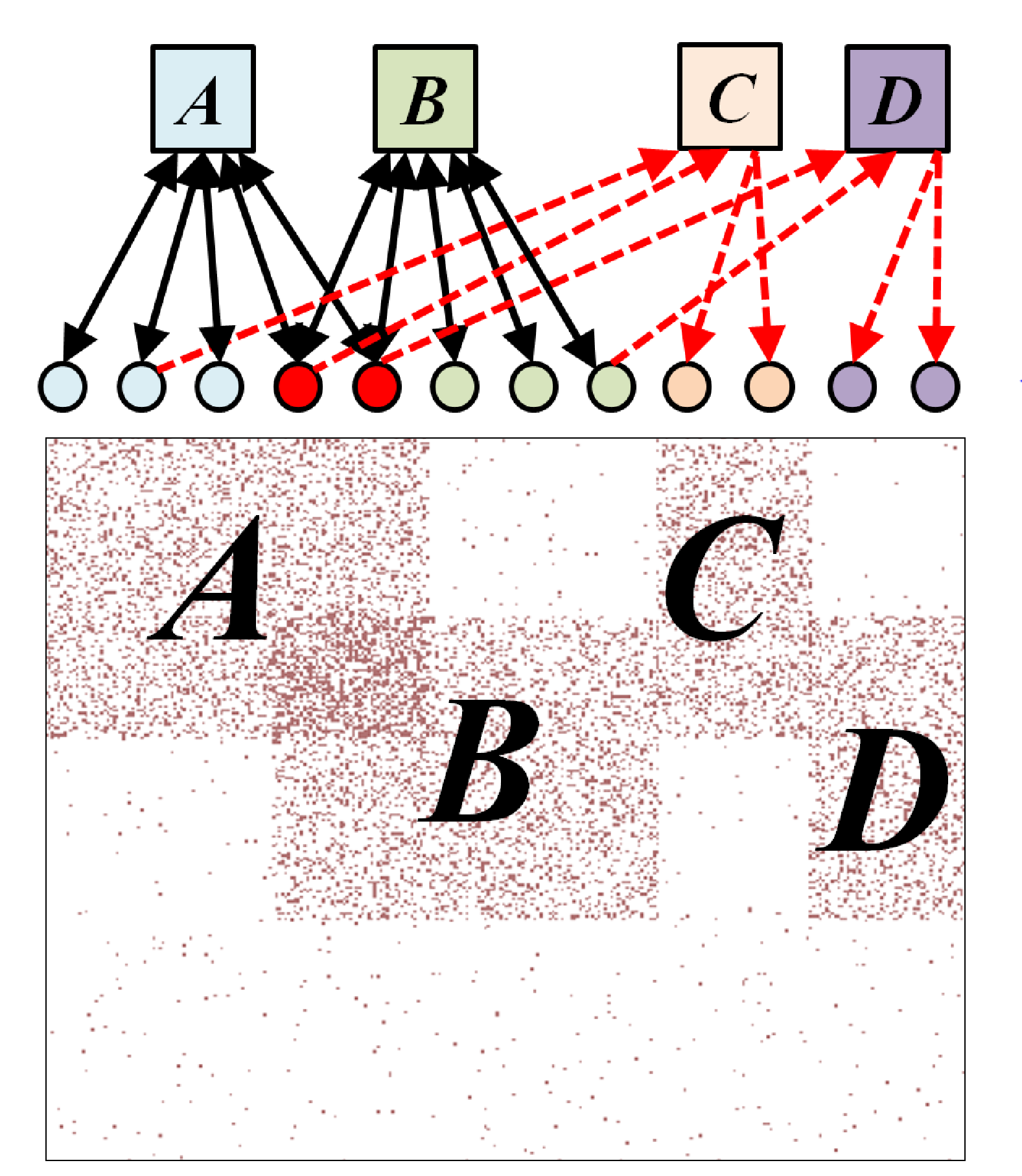}
\end{center}
\vspace{-4mm}
\caption{Affiliation graph (top) of the Directed Affilation Network Model that corresponds to the network adjacency matrix (bottom). It contains two overlapping cohesive ($A$, $B$) and two overlapping 2-mode ($C$, $D$) communities. Black edges in the affiliation graph denote bidirectional community memberships and red edges denote unidirectional memberships.}
\vspace{-2mm}
\label{fig:adjacency}
\end{figure}



Our Directed Affiliation Network Model and the underlying generated network are illustrated in Figure~\ref{fig:adjacency}. Directed affiliations are able to explain the overlapping nature of cohesive as well as 2-mode communities. For example, imagine a Twitter network among a community of music fans ($A$), a community of movie fans ($B$), a group of famous singers ($C$), and a group of famous actors ($D$). Members in communities $A$ and $B$ build bi-directional social relationships inside their respective communities. Some nodes may belong to both communities $A$ and $B$ as they are interested in both movies and music. As for one-directional relationships, we can easily see that music fans would follow singers ($C$) and movie fans would follow actors ($D$). Together, these relations would form the adjacency matrix at the bottom of Figure~\ref{fig:adjacency}. Our model captures this complex community structure very naturally, as shown in the community affiliation graph above the adjacency matrix, where green nodes represent music fans ($A$), blue nodes are movie fans ($B$), red nodes are fans of both movies and music, ivory nodes are singers ($C$), and purple nodes are actors ($D$). Affiliations between nodes and cohesive communities $A$ and $B$ flow in both directions because members of those communities have reciprocal relationships with each other, whereas fans and celebrities belonging to 2-mode communities $C$ and $D$ have edges flowing in only one direction (fans follow celebrities, celebrities are followed by fans).

More generally, our model has two important advantages over existing approaches~\cite{Lattanzi09AffiliationNetworks,jaewon12agmfit,jaewon13agmfast,zheleva09affiliation}:
First, \model can model natural overlaps between communities. It has been shown that community affiliation models for undirected networks~\cite{jaewon12agmfit} can model community overlaps accurately, which traditional models of overlapping communities fail to capture~\cite{Ahn10LinkCommunitiesNature,airoldi07blockmodel,palla05_OveralpNature}. The model also captures realistic community overlaps because its modeling power \emph{generalizes} that of other community affiliation models for undirected networks, \ie, \model can model overlaps between cohesive communities \emph{in addition to} 2-mode communities.
%
The second advantage of our model is its ability to model 2-mode communities. By modeling such communities, we can better capture the interaction between groups of nodes. This is a significant improvement over current methods that model only interactions \emph{within} communities.


\hide{ 
\xhdr{Motivation}
Our goal is to analyze how communities in directed networks differ from those in undirected networks.
To achieve this, we analyze hand-labeled communities in Twitter, Google+, and Facebook, collected by McAuley and Leskovec~\cite{nips2012}.
In Facebook, users construct `lists' to group related sets of friends, while In Twitter and Google+---where edges encode \emph{following} relationships---users create `lists' (Twitter) or `circles' (Google+) from sets of users whom the follow.

In \cite{nips2012}, the authors collected hand-labeled lists and circles, and formulated the task of discovering them as a community detection problem on the underlying ego-networks of each user, \ie, on the graph formed by friends who are mutual friends of each other. In total, this dataset consists of 1,143 ego-networks, containing 192,075 users in 5,541 circles. Table~\ref{tab:data} describes the statistics of this dataset.

\begin{table}[t]
\centering
\setlength{\tabcolsep}{3pt}
\small
    \begin{tabular}{l||r|r|r|r|r|r|r}
    \toprule
    Dataset & $N$& $E$ & $C$ & $B$ & $B_C$ & $\mathit{CCF}$ & $\mathit{FT}$ \\ \hline \hline
    \midrule
    Facebook&4,039&88,234 & 193&  1.0 & 1.0 & 0.56 & 0.66\\ \hline
    Twitter& 81,306 & 1,768,149 & 4,869 & 0.54 & 0.67 & 0.53 & 0.69 \\ \hline
    Google+&107,614& 13,673,453& 479 & 0.29& 0.36 & 0.51 & 0.38 \\
    \bottomrule
    \end{tabular}
    \caption{Dataset statistics. $N$: number of nodes, $E$: number of edges, $C$: number of communities, $B$: Fraction of bi-directional edges in the networks, $B_C$ Fraction of bi-directional edges in communities, $\mathit{CCF}$: Clustering coefficient, $\mathit{FT}$: Transitivity between the two community members followed by the same user.}
    \label{tab:data}
\setlength{\tabcolsep}{6pt}
\end{table}

In \cite{nips2012}, it was discussed how the nature of relationships in Facebook is different than those in Google+ and Twitter (due to directedness), and how this may change the interpretation of a `community' in the two networks.

To evaluate this claim, we first measure the extent to which relationships are asymmetric in directed networks. In particular, we measure the ratio $B$ of pairs of nodes whose edges flow in both direction versus the number of all connected pairs of nodes. This value is low for directed networks: 0.29 in Google+ and 0.54 in Twitter (it is 1.0 in Facebook by definition, since edges are undirected). Even when considering pairs of nodes who belong to the same groundtruth community, we find that this value $B_C$ is only slightly higher: 0.36 in Google+ and 0.67 in Twitter. Thus many communities in such networks are \emph{not} formed by highly connected sets of mutual friends.

We next examine the extent to which communities are `bipartite-like' (as in Figure~\ref{fig:diag_bipartite}). To measure this, we examine the transitivity $\mathit{FT}$ between two community member nodes who are followed by the same node outside the community, \ie, the fraction of closed triads between two members and one non-member where the two members have incoming edges from the non-member (red edge in Figure~\ref{fig:transitivity}). We then compare this value to the global transitivity in the graph (Clustering coefficient~\cite{watts98collective}). If the community has a bipartite structure such as in Figure~\ref{fig:diag_bipartite}, then the value of $FT$ would be lower than the clustering coefficient. Google+ communities have very low transitivity (0.38) compared to the clustering coefficient (0.51).
%


\begin{figure}[t]
\centering
  \subfigure[Community Affiliations]{\label{fig:dagm.bipartite}\parbox{40mm}{\centering\includegraphics[height=0.12\textwidth]{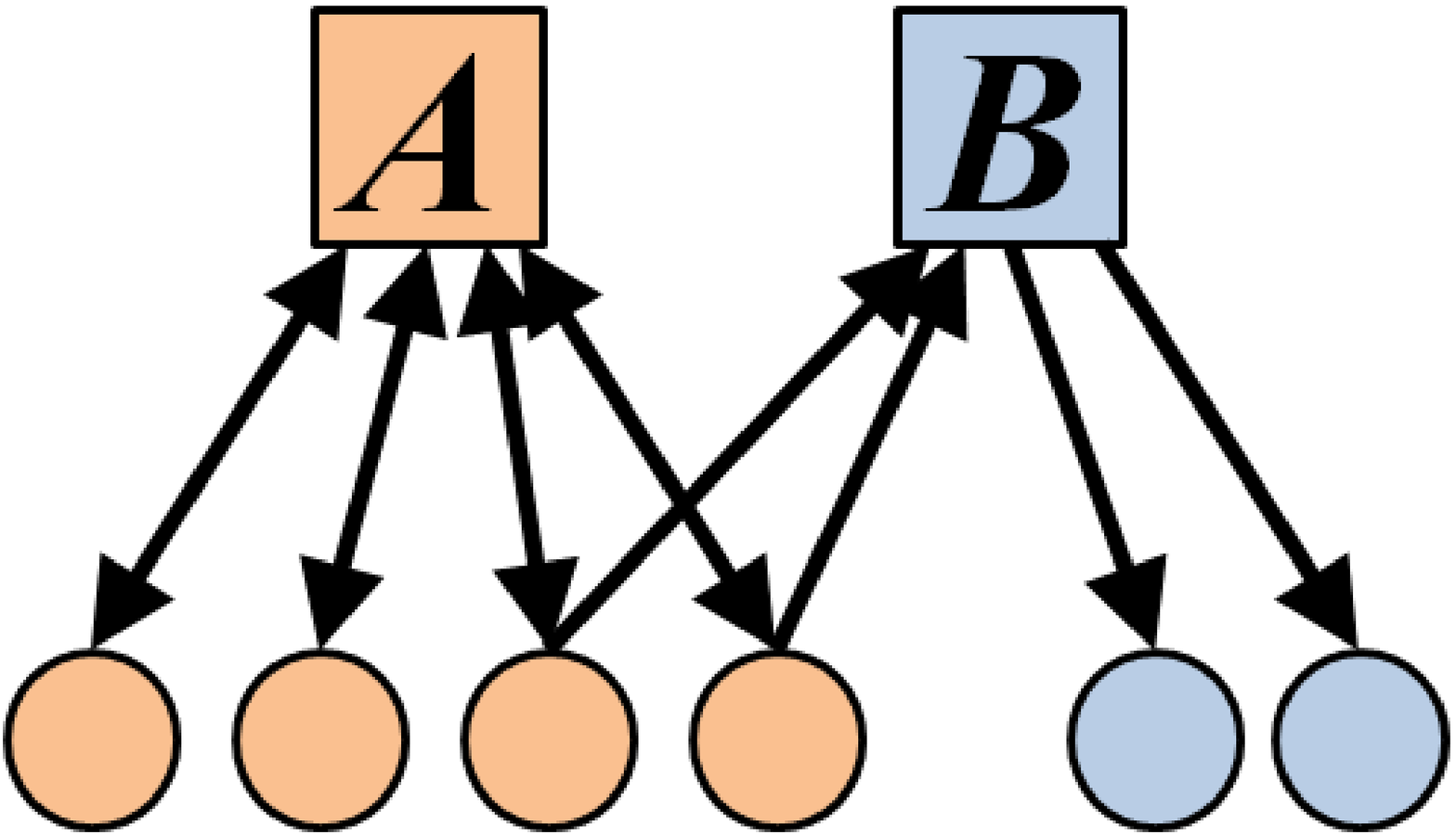}}}%
  \subfigure[Network]{\label{fig:dagm.network}\parbox{40mm}{\centering\includegraphics[height=0.12\textwidth]{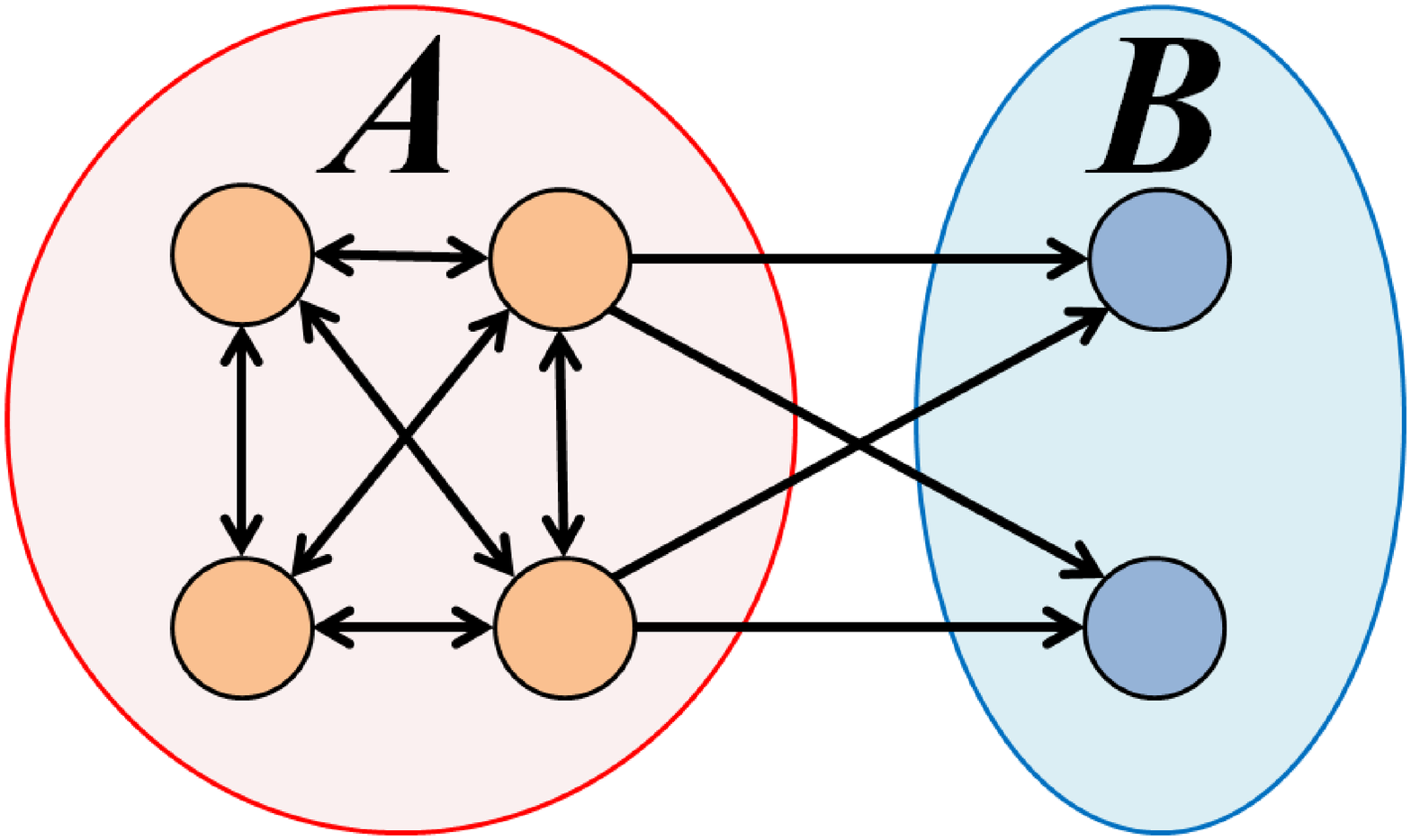}}}
  \caption{
  (a) Bipartite community affiliation graph. Squares: communities, Circles: nodes in (b). Edges from nodes to communities indicate that nodes have \emph{outgoing} edges to other members in those communities. Edges from communities to nodes indicate that nodes have \emph{incoming} edges from other members.
  (b) Network generated by \model.
  }
  \label{fig:network.illustration}
\end{figure}


These observations suggest that a desirable model of communities in directed networks should account for two properties not required when modeling undirected networks. First, the model should account for asymmetries in the relationships between the nodes. Second, communities should be modeled not only in terms of the internal density of member nodes, but also in terms of non-members to whom member nodes are connected. Applying undirected models to directed networks \emph{without} accounting for such properties is unlikely to yield good results.

Furthermore, we will demonstrate that accounting for these properties may help even when modeling communities in \emph{undirected} networks. Even when edge directions are not explicit, relationships may be asymmetric, and communities may be formed by sparsely-connected sets of nodes with common external connections.


\xhdr{\fullmodel}
We present \fullmodel (\model), a novel overlapping community detection method for directed networks.
First we present a stochastic generative model of directed networks in which the probability of an edge appearing is a function of its community affiliations. We then proceed to develop an efficient fitting procedure which allows us to detect community affiliations of nodes in a given unlabeled network.

The intuition behind our model is based on Breiger's foundational work \cite{breiger74groups} which recognized that shared group affiliations lead to communities~\cite{breiger74groups,feld86focused,simmel64affiliations}.
To represent node community memberships, we consider a bipartite affiliation graph where the nodes in the network are connected to the communities to which they belong.



In undirected networks, when a node belongs to a community it typically means that they have (undirected) relationships to other members of the community. In \cite{jaewon13agmtist}, this type of community affiliation is modeled using a bipartite graph between nodes and communities: undirected edges are formed between nodes and communities to denote community affiliation. In directed networks, however, we need a richer notion of community affiliation: a node may have many edges that go \emph{to} members of a community, many edges \emph{from} members of a community, or both.
Therefore, we assume that nodes can have two `types' of community affiliation. To model this, we assume that the membership between nodes and communities \emph{also} has a direction: outgoing membership and incoming membership.

Formally, we represent nodes' community membership using a bipartite graph as in Figure~\ref{fig:dagm.bipartite}. Round nodes (Figure~\ref{fig:dagm.bipartite}, a) represent the nodes of the social network, while square nodes represent communities. Edges indicate node community memberships. Edges have direction: Outgoing edges from nodes to communities mean that the nodes have outgoing edges to other member nodes, and incoming edges from communities to nodes mean that the nodes have incoming edges from other member nodes. Thus the community affiliation graph in Figure \ref{fig:dagm.bipartite} (a) models the directed network depicted in Figure \ref{fig:dagm.bipartite} (b). We denote the bipartite affiliation network as $B(V,C,M)$, where $V$ is the set of nodes of the underlying network $G$, $C$ the set of communities, and $M$ the directed edge set. An outgoing membership of a node $u \in V$ to a community $c \in C$ means $(u, c) \in M$ and, and an incoming membership means $(c, u) \in M$.

Now, given the affiliation network $B(V,C,M)$, we model a generative process for a network $G(V,E)$. To achieve this, we need to specify the process that generates the edges $E$ of $G$ given the affiliation network $B$. We consider a simple parameterization where we assign a parameter $p_c$ to every community $c \in C$. The parameter $p_c$ models the probability of an edge forming from a member node $u$ with \emph{outgoing} membership to community $c$ to another member node $v$ with \emph{incoming} membership. In other words, we simply generate an edge between a pair of nodes who have a path in the bipartite graph $B$ via community $c$ with probability $p_c$. Each community $c$ creates edges independently. However, if two nodes are connected by more than one community, duplicate edges are not included in the graph $G(V,E)$.

\begin{definition} Let $B(V,C,M)$ be a bipartite graph where $V$ is a set of nodes, $C$ is a set of communities, and an edge $(u,c) \in M$ means that node $u \in V$ belongs to community $c \in C$ in an outgoing way and $(c, u) \in M$ means that $u$ belongs to $c$ in an incoming way. Let also $\{p_c\}$ be a set of probabilities for all $c \in C$. Given $B(V,C,M)$ and $\{p_c\}$, \fullmodel generates a graph $G(V,E)$ by creating directed edges $(u,v)$ from nodes $u \in V$ to nodes $v \in V$ with probability $p(u,v)$:
\beq
p(u,v)= 1 - \prod_{k \in C_{uv}} (1 - p_k),
\label{eq:puv}
\eeq
where $C_{uv} \subset C$ is a set of communities through which $u$ has a path to $v$ ($C_{uv} = \{c | (u,c), (c, v) \in M\}$).
\end{definition}

\begin{figure}[t]
\begin{center}
 \includegraphics[width=0.5\linewidth]{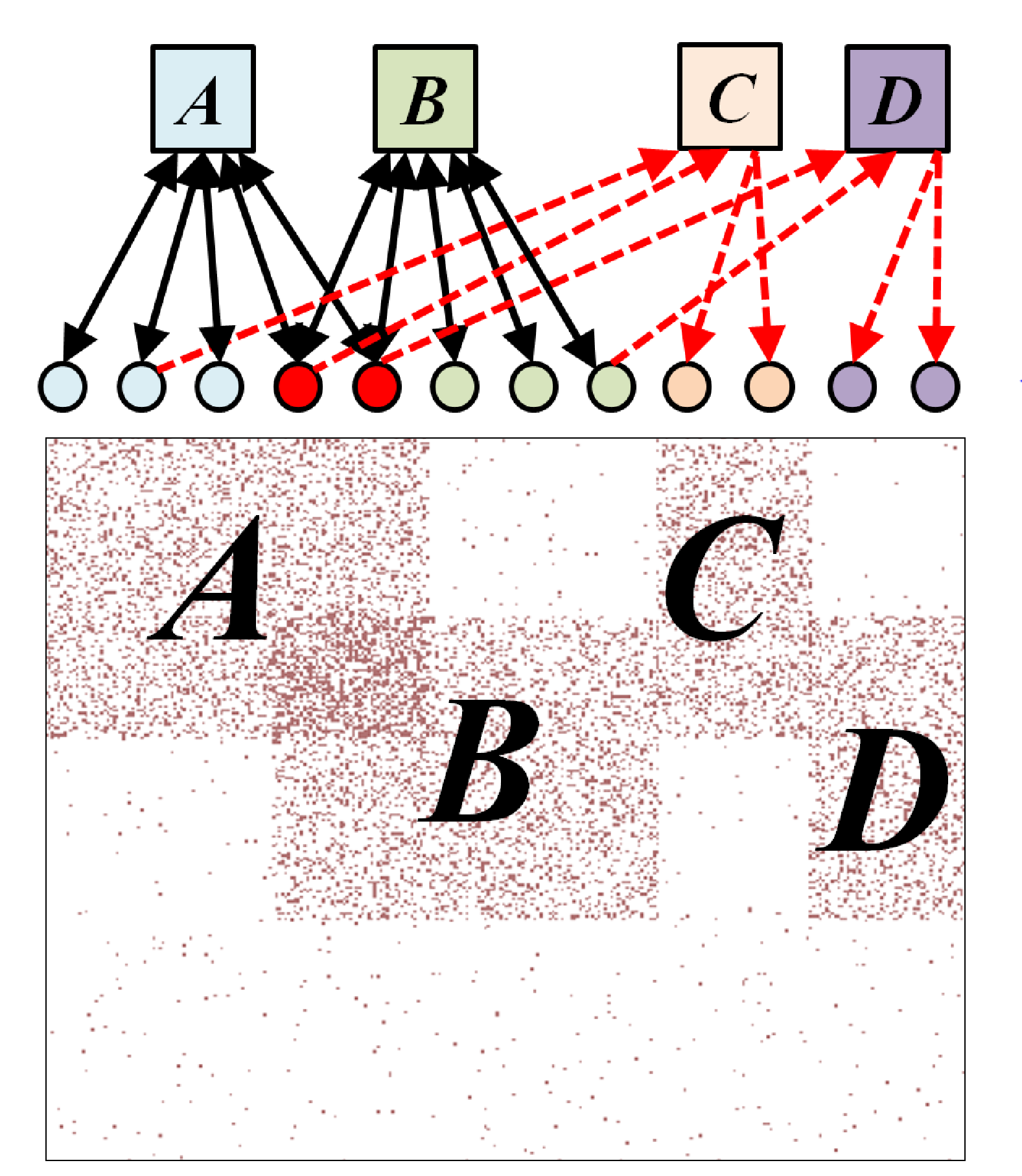}
\end{center}
\label{fig:adjacency}
\vspace{-4mm}
\caption{$A$, $B$: Two overlapping cohesive communities. $C$, $D$:  Each of the two cohesive communities follows (has outgoing edges to) different sets of other nodes. Top: community affiliation. Black edges mean memberships in both direction and red edges mean memberships in only one direction.}
\vspace{-4mm}
\end{figure}

Our model has two significant advantages compared to existing community affiliation based models. First, \model can model natural overlaps between communities. It has been shown that community affiliation models for undirected networks~\cite{jaewon12agmfit} can capture community overlaps accurately, which traditional models of overlapping communities fail to discover~\cite{Ahn10LinkCommunitiesNature,airoldi07blockmodel,palla05_OveralpNature}. \model also captures the realistic community overlaps because its modeling power \emph{generalizes} that of other community affiliation models for undirected networks, \ie, \model can model overlaps between cohesive communities \emph{in addition to} bipartite communities.

The second advantage of \model is its ability to model bipartite communities. By modeling such communities, \model can capture the interaction between the groups of nodes. This is a significant improvement over community affiliation models that model only the interaction \emph{within} communities.

These two advantages are illustrated in Figure~\ref{fig:adjacency}. Imagine a Twitter network with four communities: music fans ($A$), movie fans ($B$), famous singers ($C$), and famous actors ($D$). Members in communities $A$ and $B$ create bi-directional social relationships inside their respective communities. Some nodes may belong to the both communities $A$ and $B$ simultaneously as they are interested in both movies and music. As for unidirectional relationships, we observe that music fans follow singers ($C$) and movie fans follow actors ($D$). Together, these relations form the adjacency matrix illustrated at the bottom of Figure~\ref{fig:adjacency}. Our model can model this complex community structure naturally using the community affiliation graph, where green nodes represent music fans ($A$), blue nodes are movie fans ($B$), red nodes are fans for both movies and music, ivory nodes are singers ($C$), and purple nodes are actors ($D$). Edges between nodes and cohesive communities $A$ and $B$ flow in both directions because members of those communities have reciprocal relationships with each other, whereas fans and celebrities belonging to 2-mode communities $C$ and $D$ and have edges flowing in only one direction (fans follow celebrities, celebrities are followed by fans).



} 

\vfill\eject

\section{Community detection}
\label{sec:fitting}

Given an unlabeled, directed network $G(V, E)$, our goal is to identify cohesive as well as 2-mode communities. We achieve this by fitting our {\em Directed Affiliation Network Model} to $G(V,E)$, \ie, by finding an affiliation graph $B$ and parameters $\{p_c\}$ that maximize the data likelihood.
For now, we assume that the number of communities $K$ is given; we will later discuss how to automatically determine $K$. We aim to solve the following Maximum Likelihood Estimation problem:
\beq
\argmax_{P, \{p_c\}} \sum_{(u, v) \in E} \log p(u,v) + \sum_{(u, v) \not\in E} \log (1-p(u,v)),
\label{eq:mle_agm}
\eeq
where the edge probability $p(u,v)$ is defined in Eq.~\ref{eq:puv}.

Eq.~\ref{eq:mle_agm} leads to a challenging optimization problem. Specifically, it involves a combinatorial search over all possible affiliation graphs $B$~\cite{jaewon12agmfit}. Therefore, we develop an approximate algorithm for optimizing Eq.~\ref{eq:mle_agm}. We achive this by relaxing the original problem by changing binary memberships into real-valued memberships.


We build on the intuition from the BigCLAM~\cite{jaewon13agmfast} optimization procedure and begin by introducing variables to represent the memberships of the nodes. As noted earlier, we distinguish nodes' incoming memberships and outgoing memberships.
In particular, let $M_{uc}$ indicate whether the node $u$ belongs to community $c$ with an outgoing membership, and $L_{vc}$ indicate whether node $v$ has an incoming membership for $c$.
Now Eq.~\ref{eq:puv} can be represented as:
\[
p(u,v)= 1 - \prod_{c \in C_{uv}} (1- p_c) = 1 - \prod_{c}(1- p_c) ^ {M_{uc} L_{vc}},
\]
By applying the change of variables $1 - p_c = \exp(- \alpha_c)$ with $\alpha_c \geq 0$, the equation becomes linear in $M$, $L$, and $\alpha_c$:
\[
p(u,v)= 1 - \exp({- \sum_c M_{uc} \alpha_c L_{vc}}).
\]
We then further simplify the equation by letting $\tilde{M}_{uc} = \sqrt{\alpha_c} M_{uc}$ and $\tilde{L}_{vc} = \sqrt{\alpha_c} L_{vc}$.
\[
p(u,v)= 1 - \exp({- \sum_c \tilde{M}_{uc} \tilde{L}_{vc}}).
\]

So far, we have not used any approximations and the problem is still combinatorial since the variables remain restricted: $\tilde{M}_{uc} \in \{\sqrt{\alpha_c}, 0\}$ and $\tilde{L}_{vc} \in \{\sqrt{\alpha_c}, 0\}$.

However, note that we can interpret $\tilde{M}_{uc}$ as the strength of the membership of node $u$ to community $c$. Thus the condition $\tilde{M}_{uc} \in \{\sqrt{\alpha_c}, 0\}$ simply means that if node $u$ belongs to $c$, it would be connected to other member nodes in $c$ with the factor $\sqrt{\alpha_c}$, which determines $p_c$. The same argument also applies to $\tilde{L}_{vc}$.
%

Now we replace  $\tilde{M}_{uc}$ and $\tilde{L}_{vc}$ with {\em nonnegative continuous} valued memberships $F_{uc}$ and $H_{vc}$, respectively. The advantage here is that now each node can pick the ``strength'' of its membership to a given community: A high value of $F_{uc}$ means that the node $u$ has many outgoing edges towards other members of $c$, while high $H_{vc}$ means that node $v$ has many incoming edges from other members of $c$. Now we ca write:
\[
p(u,v)= 1 - \exp( {- F_{u}  H_{v}^T}).
\]
And we transformed Eq.~\ref{eq:mle_agm} into a continuous optimization problem:
\beq
\{\hat{F}, \hat{H}\} = \argmax_{F, H \geq 0} l(F, H)
\label{eq:mle_fgm}
\eeq
where
\[
l(F, H) = \sum_{(u, v) \in E} \log (1 - \exp( - F_{u} H_{v}^T)) - \sum_{(u, v) \not\in E} F_{u} H_{v}^T.
\]
In other words, in order to detect network communities we fit our model by estimating non-negative affiliation matrices $\hat{F}, \hat{H} \in \mathbb{R}^{N \times K}$ that maximize the likelihood $l(F, H) = \log P(G|F, H)$.

\xhdr{Solving the optimization problem}
To solve the problem in Eq.~\ref{eq:mle_fgm}, we adopt a block coordinate ascent approach: We update $F_u$ for each $u$ with $H$ fixed and update $H_v$ for each $v$ with $F$ fixed, \ie, we update either incoming or outgoing memberships of one node while fixing the other type of memberships. This approach has the advantage that each subproblem of updating $F_u$ and $H_v$ is convex. For brevity we describe only how to update $F_u$. Updating $H_v$ is analogous. For each $u$ we solve:
\beq
\argmax_{F_{uc} \geq 0} l(F_u),
\label{eq:mle_onenode}
\eeq
where
\[
l(F_u) = \sum_{v \in \mathcal{N}(u)} \log (1 - \exp( - F_{u} H_{v}^T)) - \sum_{v \not\in \mathcal{N}(u)} F_{u} H_{v}^T,
\]
where $\mathcal{N}(u)$ is a set of neighbors of $u$.
To solve this convex problem, we use projected gradient ascent with the following gradient:
\[
\nabla l(F_{u}) = \sum_{v \in \mathcal{N}(u)} H_{v} \frac { \exp( - F_{u} H_{v}^T)} {1 - \exp( - F_{u} H_{v}^T)} - \sum_{v \not\in \mathcal{N}(u)} H_{v}
\]
We compute the step size using backtracking line search. After each update, we project $F_u$ into a space of nonnegative vectors by setting $F_{uc} = \max (F_{uc}, 0)$.

Naive computation of $\nabla l(F_u)$ takes time $O(|V|)$. However, we reduce the computational complexity to the \emph{degree} of $u$, $O(|\mathcal{N}(u)|)$, which significantly increases the scalability of our approach.
We achieve this by computing the second term $\sum_{v \not\in \mathcal{N}(u)} H_{v}$ in $O(|\mathcal{N}(u)|)$ by storing/caching $\sum_v H_{v}$:
\[
\sum_{v \not\in \mathcal{N}(u)} H_{v} = (\sum_v H_{v} - H_u - \sum_{v \in \mathcal{N}(u)} H_{v}).
\]
Given that real-world networks are extremely sparse ($|\mathcal{N}(u)| \ll N$), we can update $F_u$ for a single node $u$ in \emph{near-constant} time. The update rule for $H_v$ can be similarly derived and takes near-constant time $O(|\mathcal{N}(v)|)$. In practice, we iteratively update $F_u$, $H_u$ for each $u$ and stop iterating once the likelihood does not increase (by $0.01\%$) after we update $F_u$, $H_u$ for all $u$.

\xhdr{Determining community affiliations of nodes}
From the real-val\-ued $\hat{F}, \hat{H}$ that we estimate, we want to determine ``hard'' community affiliations of nodes. We achieve this by thresholding $F_{uc}$ and $H_{uc}$ with a constant $\delta$, \ie, we regard $u$ has an outgoing membership to community $c$ if $F_{uc} \geq \delta$, and an incoming membership from $c$ if $H_{uc} \geq \delta$.

We choose the value of $\delta$ so that every pair of members in community $c$ has edge probability higher than the background edge probability $1 / |V|$ (see Eq.~\ref{eq:puv}):
\[
\frac{1}{|V|} \leq 1 - \exp(- \delta^2)
\]
This inequality leads to $\delta = \sqrt{- \log (1 - 1 / |V|)}$. We note that we also experimented with other values of $\delta$ and found that this choice for $\delta$ works well in practice.


\xhdr{Algorithm initialization}
To initialize $F, H$, we employ \emph{locally minimal neighborhoods}, which provide good seed-sets for community discovery~\cite{Gleich12neighborhoods}. A neighborhood $N(u)$ of a node $u$ is a set consisting of the node $u$ and its neighbors, and $N(u)$ is said to be ``locally minimal'' if $N(u)$ has lower conductance score than $N(v)$ for any other neighbor $v$ of $u$~\cite{Gleich12neighborhoods}. For a node $u'$ belonging to such a locally minimal neighborhood $k$, we initialize $F_{u'k} = 1$ if $u'$ has an outgoing edge (or $F_{u'k} = 0$ otherwise), and set $H_{u'k} = 1$ if $u'$ has an incoming edge (or $H_{u'k} = 0$ otherwise).

\xhdr{Choosing the number of communities}
To automatically determine the number of communities $K$, we follow the approach proposed in \cite{airoldi07blockmodel}. We divide all node pairs into 80\% training and 20\% test set. Varying $K$, we fit \model with $K$ communities on the training pairs and measure the likelihood for the test pairs. We then select $K$ with the highest test set likelihood. For a small networks with fewer than 100 edges, we find that a different criterion works better in practice. Here we choose $K$ so as to achieve the smallest value of the Bayesian Information Criterion:
\[
\mathit{BIC}(K) = - 2 l(\hat{F}, \hat{H}) + NK\log |E|.
\]

\xhdr{Parallelization and implementation details}
Our approach also naturally allows for \emph{parallelization}, which further increases scalability of \model.
When updating $F_u$ for each node $u$ (Eq.~\ref{eq:mle_onenode}), we observe that each subproblem is {\em separable} since all other variables in Eq.~\ref{eq:mle_onenode} ($H$) remain fixed. That is, updating the value of $F_u$ for a specific node $u$ does not affect updates of $F_v$ for all other nodes $v$. In the parallelized version of \model, we solve Eq.~\ref{eq:mle_onenode} for multiple nodes in parallel. This parallelization does not affect the final result of the method. Updating $H_u$ for each node $u$ can be parallelized in the same way.
As we show in Section~\ref{sec:experiments}, parallelization on a single shared memory machine boosts the speed of \model by a factor of 20 (the number of threads) used when analyzing a 300,000 node network.
Last, we also experimented with other optimization techniques such as the cyclic coordinate descent method (CCD)~\cite{Hsieh11FastNMF} which optimizes $F_{uc}$ for each $u$ and each $c$ by Newton's method, but we found that block coordinate ascent converges the fastest.

A parallel C++ implementation of \model is publicly available at \url{http://snap.stanford.edu}.


\xhdr{\model for undirected networks}
So far, we have discussed \model under the context of directed networks. However, \model can easily be applied to undirected networks as well. We make a simple observation: undirected networks model symmetric relationships and thus an undirected relationship is equivalent to two  directed relationships, one each way. Thus, given an undirected network, we simply convert the network into a directed one by regarding every edge as reciprocal, and then apply \model to detect communities.

Now, \model will easily detect cohesive communities in this converted network as edges in cohesive communities are reciprocal.
Detecting 2-mode communities is also simple. Consider the case where we are given an undirected 2-mode community $X$ where nodes in group $A$ are connected to nodes in group $B$.
Once we convert $X$ into a directed network with reciprocal edges between $A$ and $B$, \model will estimate two 2-mode communities from this community $X$: $\hat{X}_1$  for edges from $A$ to $B$, and $\hat{X}_2$  for edges from $B$ to $A$.
Thus, \model is able to correctly discover $X$, with the caveat that it discovers it twice (both $\hat{X}_1$ and $\hat{X}_2$ correspond to $X$). 

\section{Experiments}
\label{sec:experiments}

We evaluate the performance of \model and compare it to state-of-the-art community detection methods on a range of directed as well as undirected networks. We measure the quality of community detection by computing the detection accuracy based on gold-standard ground-truth communities. We also evaluate the scalability of the methods by measuring runtime as network size increases.

\begin{table}[t]
\centering
\setlength{\tabcolsep}{3pt}
    \small
    \begin{tabular}{lcrrrrr}
    \toprule
    Dataset & Directed& $N$ & $E$ & $C$ & $S$ & $A$ \\ 
    \midrule
    Google+ & \cmark &250,469 & 30,230,905 & 437 & 143.51 & 0.25\\ 
    Twitter & \cmark & 125,120 & 2,248,406 & 3,140 & 15.54 & 0.39\\ 
    Facebook & \xmark&4,089 & 170,174 & 193 & 28,76 & 1.36\\ 
    Enron & \cmark & 45,266 &   185,172 &   4,572 & 63.93 & 6.46 \\ 
    LiveJournal& \xmark &3,997,962&34,681,189&287,512&22.31&1.59\\ 
    Youtube& \xmark& 1,134,890 & 2,987,624 & 8,385 & 13.50 &0.10\\ 
    \bottomrule
    \end{tabular}
    \vspace{-1mm}
    \caption{Dataset statistics. Directed: Yes/no, $N$: number of nodes, $E$: number of edges, $C$: number of ground-truth communities, $S$: average ground-truth community size, $A$: ground-truth community memberships per node. Further datasets used in this study are described in Table~\ref{tab:data-nogt}.
    }
    \label{tab:data-gt}
  \vspace{-3mm}
\end{table}

\subsection{Dataset Description}
We begin by briefly describing the networks that we consider in this study. Overall, we consider 5 undirected and 9 directed networks from a wide spectrum of domains. We consider social, communication, information, biological and ecological networks.\footnote{We use the publicly available data from the Stanford Large Network Collection: \url{http://snap.stanford.edu}.}

\begin{table*}[!ht]
\begin{center}
\small
\setlength{\tabcolsep}{5pt}
\begin{tabular}{l|ccc|ccc|c}
\toprule
                             & \multicolumn{3}{c|}{$F_1$ score}               & \multicolumn{3}{c|}{Jaccard similarity} & \\
Method       & Google+ & Twitter & Facebook & Google+ & Twitter & Facebook & Average\\
\midrule
MMSB \cite{airoldi07blockmodel}                     & 0.324 (0.033) & 0.262 (0.005) & 0.374 (0.042) & 0.214 (0.026) & 0.169 (0.004) & 0.266 (0.036) & 0.268\\
Clique percolation \cite{palla05_OveralpNature}     & 0.331 (0.036) & 0.246 (0.006) & 0.429 (0.051) & 0.240 (0.032) & 0.163 (0.005) & 0.342 (0.050) & 0.292\\
Link clustering \cite{Ahn10LinkCommunitiesNature}   & 0.304 (0.016) & 0.334 (0.003) & 0.372 (0.027) & 0.226 (0.016) & \bf 0.238 (0.003)$^2$\hspace{-1.4mm} & 0.275 (0.024) & 0.291\\
BigCLAM \cite{jaewon12agmfit}  & 0.324 (0.017) & 0.344 (0.005) & 0.442 (0.042) & 0.217 (0.014) & 0.234 (0.004) & 0.325 (0.038) & 0.315\\
DEMON \cite{demon}             & 0.343 (0.029) & 0.308 (0.005) & 0.418 (0.046) & 0.255 (0.027) & 0.210 (0.005) & 0.311 (0.041) & 0.307\\
NMF \cite{Lin07gradientforNMF}             & 0.333 (0.019) & 0.318 (0.004) & 0.406 (0.038) & 0.242 (0.026) & 0.221 (0.004) & 0.301 (0.050) & 0.303\\
{\bf \model}, undirected                                  & \bf 0.414 (0.027)$^1$\hspace{-1.4mm} & \bf 0.348 (0.005)$^2$\hspace{-1.4mm} & \bf 0.470 (0.042)$^1$\hspace{-1.4mm} & \bf 0.314 (0.026)$^1$\hspace{-1.4mm} & 0.237 (0.004) & \bf 0.357 (0.039)$^1$\hspace{-1.4mm} & \bf 0.357$^2$\hspace{-1.4mm}\ \\
{\bf \model}, directed                                    & \bf 0.406 (0.025)$^2$\hspace{-1.4mm} & \bf 0.363 (0.005)$^1$\hspace{-1.4mm} & \bf 0.470 (0.042)$^1$\hspace{-1.4mm} & \bf 0.314 (0.024)$^1$\hspace{-1.4mm} & \bf 0.250 (0.004)$^1$\hspace{-1.4mm} & \bf 0.357 (0.039)$^1$\hspace{-1.4mm} & \bf 0.360$^1$\hspace{-1.4mm}\ \\
\bottomrule
\end{tabular}
\normalsize
\setlength{\tabcolsep}{6pt}
\end{center}
\vspace{-3mm}
\caption{Performance on Facebook, Google+, and Twitter. Higher is better. Standard errors are shown in parentheses. The best and second best methods are annotated as `1' and `2'. \label{tab:circles}}
\vspace{-2mm}
\end{table*}

\xhdr{Networks with ground-truth communities}
For the experiments in this section, we consider a subset of 6 publicly available networks where we have explicit \emph{ground-truth} memberships of nodes to communities~\cite{jaewon11comscore}. The availability of ground-truth allows us to quantify the quality of community detection methods \emph{quantitatively}. Table~\ref{tab:data-gt} shows the statistics of the networks and the ground-truth communities. The networks come from three different domains: The first three networks are the collection of ego-networks from online social networks of Facebook, Twitter and Google+~\cite{nips2012}, the Enron email communication network~\cite{klimt04enron}, and LiveJournal and Youtube social networks~\cite{mislove07measurement}. We describe the nature of ground-truth communities in each of these datasets in more detail later.

\subsection{Experimental Setup}
\xhdr{Baselines}
For comparison we consider the following baseline met\-hods: \emph{MMSB} (Mixed Membership Stochastic Blockmodels) \cite{airoldi07blockmodel}, which can detect both cohesive and 2-mode communities in undirected networks and is extremely slow; \emph{Clique Percolation}, \cite{palla05_OveralpNature} \emph{Link Clustering} \cite{Ahn10LinkCommunitiesNature}, \emph{BigCLAM} \cite{jaewon12agmfit,jaewon13agmfast} are state of the art overlapping cohesive community detection techniques for undirected networks; \emph{DEMON} \cite{demon} is a scalable local community detection method for directed networks; \emph{NMF} \cite{Lin07gradientforNMF} is a state-of-the-art non-negative matrix factorization approach which can be used for directed networks.
We use publicly available implementations of each of the methods.

Some methods require input parameters. MMSB and NMF requires the number of communities $K$. We use the Bayes information criterion suggested by the authors~\cite{airoldi07blockmodel} to choose $K$. DEMON requires $\varepsilon$, the threshold value for merging two communities. As there exists no standard criterion for $\varepsilon$, we set $\varepsilon$ so that DEMON detects the same number of communities as \model does.

Last, we note that the above baselines represent the current state-of-the-art in community detection. However, we also considered other baselines, including those that make use of node features \cite{hierarchical}, network topology \cite{rosvall08infomap}, or both \cite{blockLDA,nips2012}; however experiments demonstrate that none of these alternatives outperforms \model.



\xhdr{Evaluation}
To evaluate the performance of the above methods we quantify the degree of correspondence between the ground-truth and the detected communities.
To compare a set of ground-truth communities $C^*$ to a set of predicted communities $C$, we adopt an evaluation procedure previously used in~\cite{jaewon12agmfit,jaewon13agmfast}, where every detected (ground-truth) community is matched with its most similar ground-truth (detected) counterpart community: 
\begin{equation*}
 \frac{1}{2|C^*|} \sum_{C^*_i \in C^*} \max_{C_j \in C}\delta(C^*_i, C_j) + \frac{1}{2|C|} \sum_{C_j \in C} \max_{C^*_i \in C^*}\delta(C^*_i, C_j),
\end{equation*}
where $\delta(C^*_i, C_j)$ is some measure of the similarity between the communities $C^*_i$ and $C_j$. We consider two standard measures of the similarity between sets, namely the $F_1$ score and the Jaccard similarity. Thus, we obtain a value between 0 and 1, where 1 indicates perfect recovery.


\subsection{Detecting Social Circles}
First we consider the problem of discovering users' social circles~\cite{nips2012}. Circles (or `lists' in Facebook and Twitter) give users a means of categorizing their immediate neighbors, or in the case of directed networks, the users whom they follow. Thus the problem of automatically identifying users' social circles can be posed as a community detection problem on each user's ego-network \cite{nips2012}.

In Table \ref{tab:circles} we evaluate the performance of \model and baselines on social circle detection. Across all three datasets and both evaluation metrics, \model (the last row) is the best or second-best performer. On average, \model outperforms MMSB by 34\%, Clique percolation by 23\%, Link clustering by 24\%, BigCLAM by 14\%, DEMON by 17\%, and NMF by 19\%.


The 3 data sets possess very different reasons for community (\ie, social circle) formation: Facebook is an undirected network and in Facebook circles are driven by dense mutual friendships among users with homogeneous backgrounds~\cite{nips2012}; therefore, we would expect cohesive communities in Facebook. Google+ and Twitter are directed networks and as such circles are not necessarily based on friendship, because edges in these networks denote \emph{follower} relationships: The fraction of reciprocated edges is only 29\% in Google+ and 54\% in Twitter. For example, a social circle in Twitter might consist of authors who publish in the same genre, or candidates in the same election. As we will see later in Section~\ref{sec:discussion}, many social circles in Google+ and Twitter follow such 2-mode structure.

Regardless of very different nature of the data sets, \model is the best performing method in each of them. This result means that \model recovers 2-mode circles in Google+ or Twitter \emph{as well as} cohesive circles in Facebook, \ie, \model can detect \emph{both} kinds of communities more accurately than the baselines.




\xhdr{Directed vs. undirected networks}
To further examine the performance out method on directed and undirected networks we perform an experiment with the goal of understanding whether \model is still able to recover 2-mode communities even when edge directions are dropped and networks are considered as undirected. To test this, we convert the directed networks of Twitter and Google+ into undirected by removing the edge directions. Then we apply \model (\model, undirected, the second to last row in Table~\ref{tab:circles}). Surprisingly, \model achieves similar performance even without explicit edge directions in the network. Based on this evidence we conclude \model is capable of accurately finding 2-mode communities even in undirected networks.


%


\begin{table}[t]
\centering
\footnotesize
    \begin{tabular}{lccc}
    \toprule
    Method & $F_1$ score & Jaccard similarity & Average \\ 
    \midrule
    MMSB& N/A& N/A & N/A\\
    Clique percolation & N/A& N/A & N/A\\
    Link clustering&	0.195 & 0.294 & 0.245 \\
    BigCLAM& 0.478 & 0.358 & 0.418\\
    DEMON&	0.464& 0.350 & 0.407 \\ 
    NMF& N/A& N/A & N/A\\
    \model, undirected & 0.538 & 0.431 & 0.485    \\
    \model, directed&	0.617& 0.516 & 0.567 \\ 
    \bottomrule
    \end{tabular}
    \vspace{-0mm}
    \caption{Performance of recipient discovery on the Enron network. Algorithms that do not scale to the size of the dataset are labeled as ``N/A''.}
    \label{tab:email}
\end{table}


\subsection{Discovering Recipient Lists in Email Networks}

We also define a task of automatically discovering recipient lists in the the email communication network. The idea is that such lists exhibit a distinct structural pattern in the network as the recipient lists may have 2-mode community structure as a set of users who receive the same email may not necessarily email each other~\cite{emailRoles}.

We consider all Enron emails~\cite{klimt04enron} with 20 or more recipients. This gives us a set of 4,572 unique recipient lists in the Enron dataset, which we treat as ground-truth communities (Table~\ref{tab:data-gt}). Now we are given an unlabeled directed Enron email communication network, where an edge $i \rightarrow j$ means that $i$ sent at least one mail to $j$, and the goal is to discover email recipient lists.

We then apply \model as well as the baselines to this network and in Table~\ref{tab:email} we measure how accurately the communities detected by \model correspond to these ground-truth email recipient lists. We report both the $F_1$ score and Jaccard similarity (for methods that do not scale to networks of this size, we report N/A). Table~\ref{tab:email} shows that \model outperforms other methods by a significant margin. \model outperforms Link clustering by 131\%, DEMON by 39\%, and BigCLAM by 36\%.

\subsection{Experiments on Large Networks}
Last, we also examine two real-world social networks with millions of nodes in which nodes explicitly declare their community memberships~\cite{jaewon11comscore}.
We consider the LiveJournal and Youtube social networks, and regard user-created groups as ground-truth communities. We ignore groups containing fewer than 10 nodes, yielding 71,093 communities in LiveJournal and 2,078 in Youtube.


Of the baselines previously mentioned, only BigCLAM could scale to both networks and DEMON could scale to the Youtube network. Therefore, we also consider two large-scale graph partitioning methods as baselines for this experiment: Metis~\cite{karypis98metis} and Graclus~\cite{dhillon07graclus}. For all methods we set the number of communities $K$ to be the number of ground-truth communities.

Table~\ref{tab:largenets} shows the results. For this experiment we focus on the score relative to that of the worst-performing baseline in each network (so that the worst-performing baseline has a score of 100\%). We compute the relative score because the networks are only partially labeled and the overall performance is thus artificially low (as methods discover many unlabeled communities).
We find that \model outperforms its nearest competitor by 8.4\% on LiveJournal and 29\% on Youtube.

\begin{table}[t]
\centering
\footnotesize
    \begin{tabular}{lcccc}
    \toprule
    & \multicolumn{2}{c}{Relative $F_1$ score}& \multicolumn{2}{c}{Absolute $F_1$ score}\\
    Method & LiveJournal & Youtube & LiveJournal & Youtube \\ 
    \midrule
    Metis  & 100\%       & 200\%   & 0.12 & 0.028\\
    Graclus& 100\%       & 185.7\% & 0.12 & 0.026\\
    BigCLAM  & 121.0 \%         & 278.1 \% & 0.14  & 0.039   \\
    DEMON  & N/A         & 100\%   & N/A & 0.014\\
    \model & 129.4\%     & 307.1\% & 0.15 & 0.043\\ 
    \bottomrule
    \end{tabular}
    \vspace{-0mm}
    \caption{Relative accuracy (compared to the worst performing method) of detected communities on large scale social networks.
    }
    \label{tab:largenets}
\end{table}

\subsection{Scalability}

\begin{figure}[t]
\begin{center}
\includegraphics[width=0.85\linewidth]{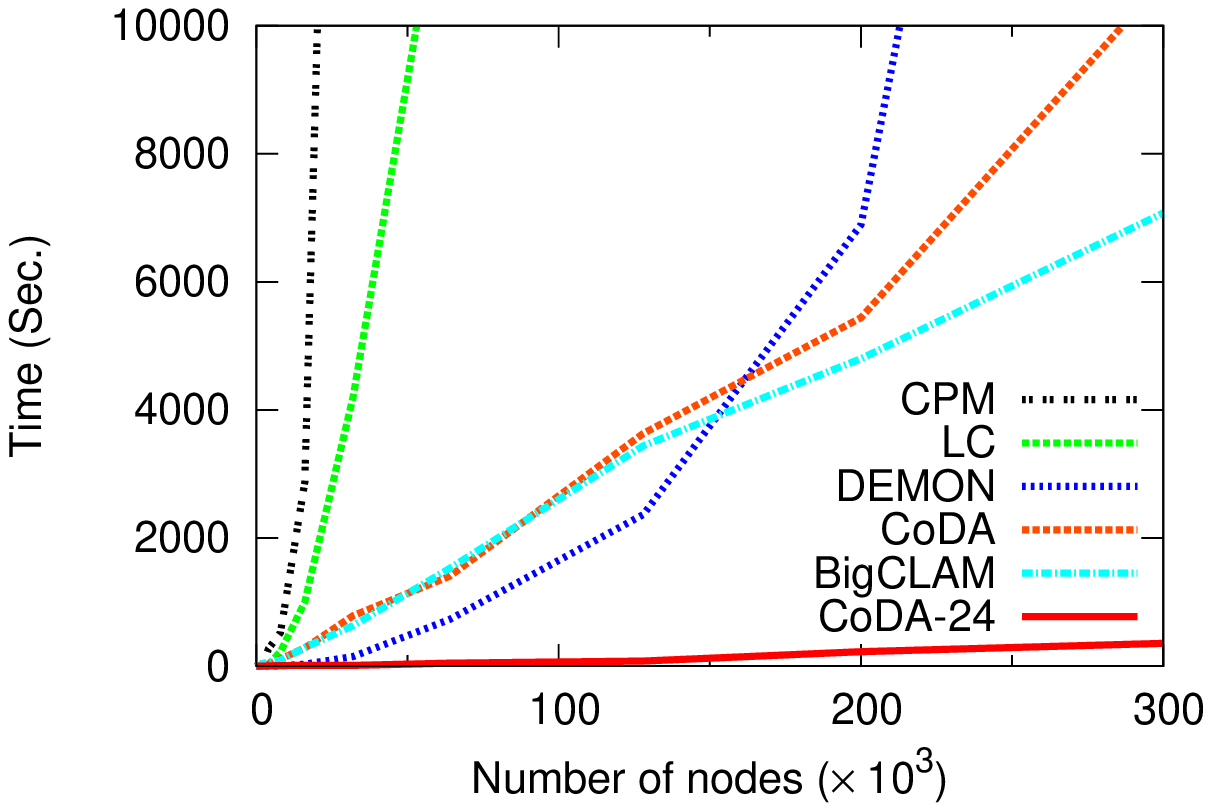}
\end{center}
\vspace{-5mm}
\caption{Algorithm runtime.
}
\vspace{-1mm}
\label{fig:scalability}
\end{figure}

Last, we evaluate the scalability of \model by measuring its running time on synthetic networks with increasing size. We generate synthetic networks using the Forest fire model~\cite{jure05dpl} with the forward and backward probabilities set to 0.36 and 0.32, respectively.
Since \model is easily parallelizable as described in Section~\ref{sec:fitting}, we also consider a singe machine parallel implementation running with 24 threads (\model-24).

Scalability results are shown in Figure~\ref{fig:scalability}. Link Clustering and Clique Percolation scale to networks of at most a few thousand nodes. DEMON is a fast and scalable overlapping community detection method. DMEON tends to be faster than \model (single-threaded implementation) for networks up to 100,000 nodes, however, once the network becomes larger, \model becomes much faster.

When comparing single-threaded implementations we also note that BigCLAM is the fastest method in our experiments. However, we note that \model takes only 30\% more time than BigCLAM while it is also solving a more complicated problem, namely detecting cohesive as well as 2-mode communities.

Last, we also measure a parallelized version of \model (\model-24). Using 24 threads on a single machine, we achieve nearly 24x speedup. Ultimately, \model takes just 6 minutes to process a 300,000 node network.



\section{Community discovery}
\label{sec:discussion}
So far we have demonstrated that \model can reliably detect both cohesive and 2-mode communities in directed as well as undirected networks. 
In the following section, we shall demonstrate that 2-mode communities
take an important role in networks.
We shall use \model to perform a qualitative study of various networks in order to determine the extent to which community structures vary across real-world networks from various domains.



\xhdr{Network data}
In addition to the datasets already introduced, we also analyze biological networks, foodwebs, web graphs, and citation networks (Table~\ref{tab:data-nogt}). For biological networks, we consider the protein-protein interaction network of \emph{Saccharomyces cerevisiae}: yeast two-hybrid (PPI-Y2H) and literature-curated (PPI-LC)~\cite{Ahn10LinkCommunitiesNature}. We also consider the Chesapeake and Florida Bay foodwebs~\cite{ulanowicz98foodweb_baywet}, the web graph of Stanford University web pages (web-Stanford), the web graph released by Google in 2002 (web-Google)~\cite{jure08ncp2}, and the arXiv citation networks from high-energy physics phenomenology (cit-HepPh) and theory (cit-HepTh)~\cite{jure05dpl} all available from \url{http://snap.stanford.edu}.

\begin{table}[t]
\centering
\setlength{\tabcolsep}{3pt}
    \begin{tabular}{lcrrrrr}
    \toprule
    Dataset & Directed& $N$ & $E$ & $C$ & $S$ & $A$ \\ 
    \midrule
    PPI-Y2H & \xmark & 1,647 & 2,518 & 40 & 90.75 & 2.20 \\
    PPI-LC & \xmark &1,213 & 2,556 & 40 & 42.08 & 1.39 \\
    web-Stanford & \cmark&281k & 2,312k & 19k & 70.63 & 4.59 \\
    web-Google & \cmark & 875k &	5,105k & 39k & 41.79 & 1.86 \\
    cit-HepTh & \cmark &27k & 353k & 2,000 & 70.00 & 5.04\\ 
    cit-HepPh  & \cmark& 34k & 422k & 4,976 & 51.52 & 7.42 \\ 
    Florida Bay& \cmark & 121& 1,745 & 6 & 45.33 & 2.25\\ 
    Chesapeake & \cmark & 33 &	72 & 5 & 9.20 & 1.39 \\
    \bottomrule
    \end{tabular}
     \vspace{-2mm}
    \caption{Dataset statistics.
    Directed: Whether the network is directed or not, $N$: number of nodes, $E$: number of edges, $C$: number of detected communities, $S$: average size of detected communities, $A$: community memberships per node.
    }
    \label{tab:data-nogt}
   \vspace{-4mm}
\end{table}


\subsection{Biological and Foodweb Communities}

\begin{figure}[t]
  \begin{center}
  \vspace{2mm}
  \includegraphics[width=0.8\linewidth]{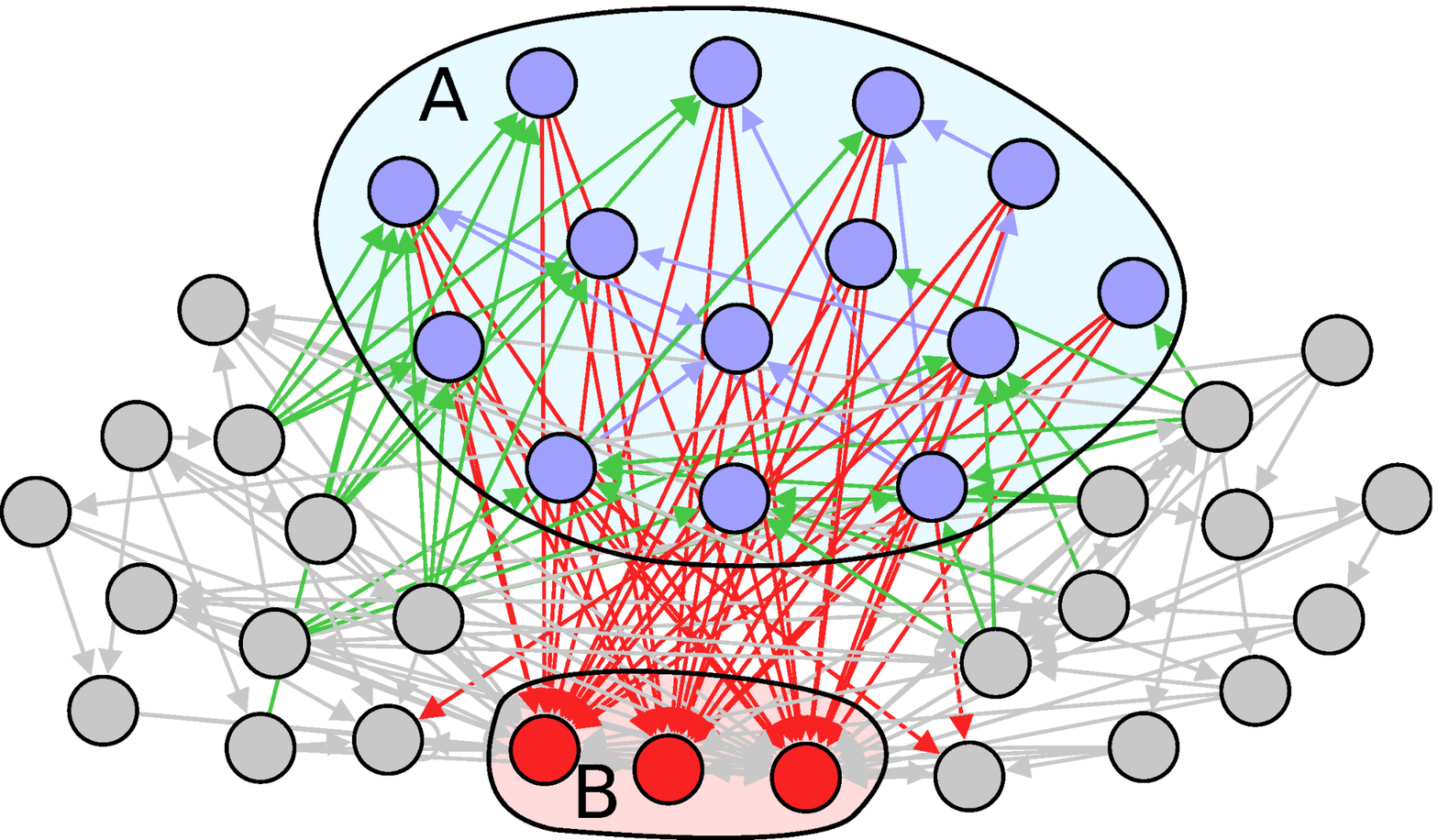}
  \end{center}
   \vspace{-5mm}
  \caption{Two detected communities in a Foodweb (Chesapeake Bay). Among other communities, \model identifies sets of nodes with similar predators ($A$, blue nodes) and with similar prey ($B$, red nodes), both of which have low internal connectivity.
  \label{fig:foodweb}}
   \vspace{-3mm}
\end{figure}

\begin{figure}[t]
\begin{center}
\includegraphics[width=1.00\linewidth]{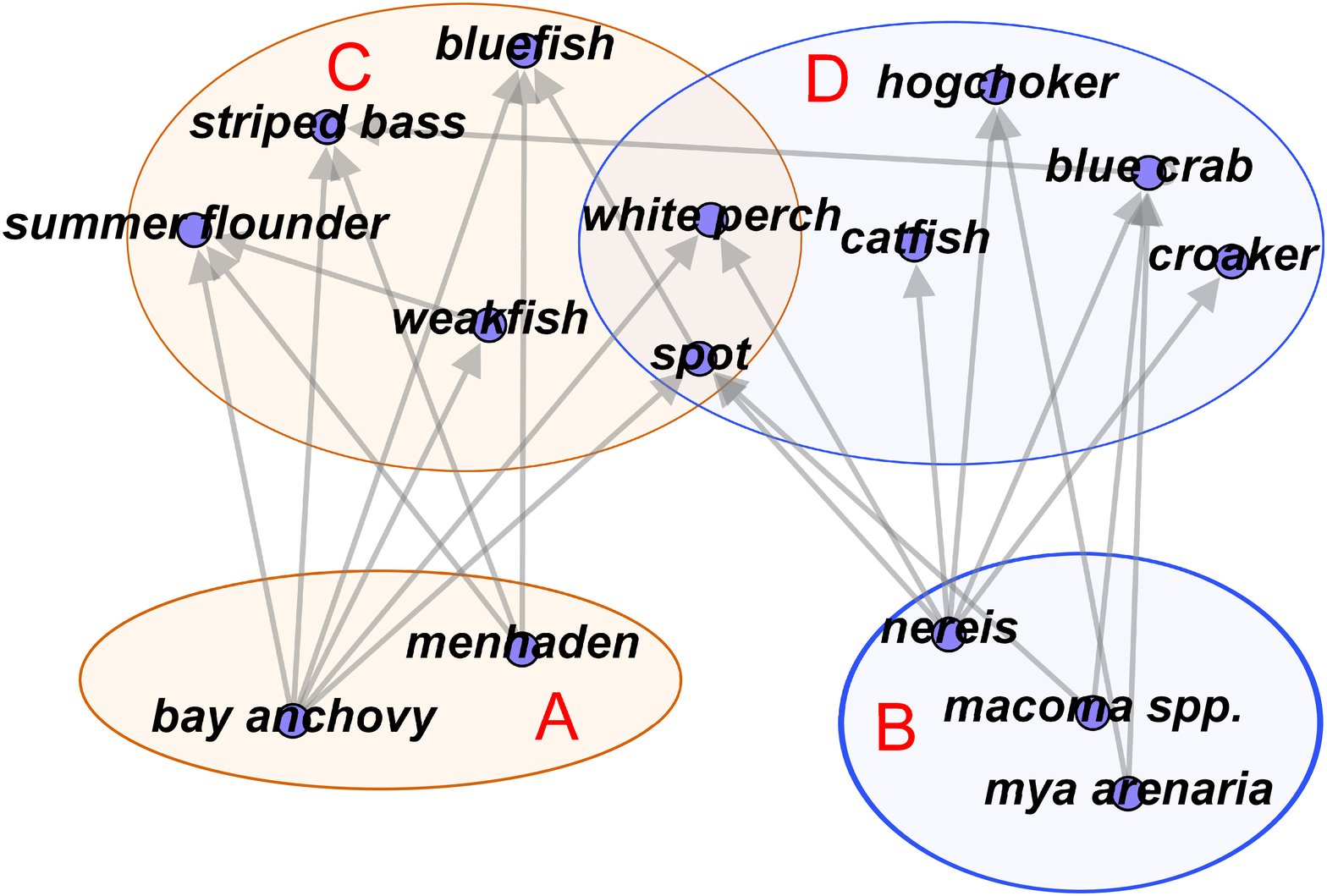}
\end{center}
 \vspace{-5mm}
\caption{Examples of overlapping 2-Mode communities detected by our method in the Chesapeake bay foodweb network. See main text for the explanation of community structure. \label{fig:foodweb_2mode}}
\end{figure}

We first present 2-mode communities in foodwebs,
where nodes represent organisms and an edge from a node $u$ to $v$ means that $u$ is preyed upon by $v$.
We apply \model on the Chesapeake Bay foodweb network shown in Figure~\ref{fig:foodweb}, and display an induced subgraph of detected 2-mode communities in Figure~\ref{fig:foodweb_2mode}.

In foodweb networks, we find 2-mode communities of groups of predators who rely on similar groups of prey (Figure~\ref{fig:foodweb_2mode}). 
The blue 2-mode community ($B$-$D$) represents predators and prey in the the Chesapeake Bay sands: \emph{nereis}, \emph{macoma spp.}, and \emph{mya arenaria} (in $B$) are small, sand-dwelling clams and worms that are fed on by fish (in $D$).
Alternately, the red community ($A$-$C$) shows predator-prey relationships among fish: small fish ($A$) are eaten by bigger fish ($C$). \model also discovers the \emph{overlap} between two predator groups where \emph{white perch} and \emph{spot} prey on both fish and clams.

\begin{figure}[t]
\begin{center}
\includegraphics[width=1.07\linewidth]{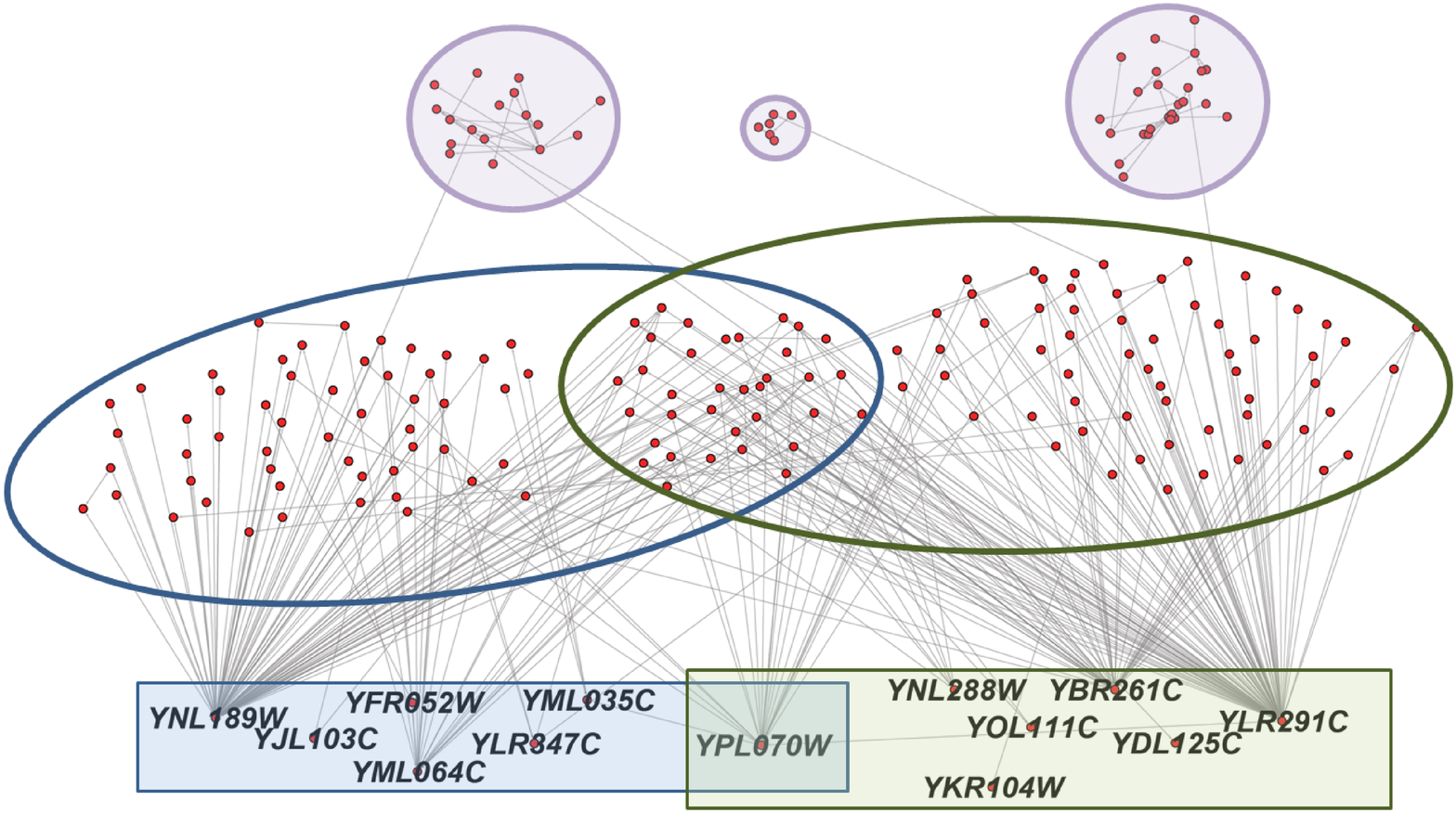}
\end{center}
\vspace{-3mm}
\caption{Overlapping 2-Mode communities detected by our method in a Protein-Protein interaction network. See main text for the explanation of community structure. \label{fig:ppi_2mode}}
\vspace{-3mm}
\end{figure}

\model also allows us to gain insights into biological PPI networks.
Interestingly, \model discovers many 2-mode communities in the undirected protein-protein interaction network determined by yeast two-hybrid screening (PPI-Y2H). For example, Figure~\ref{fig:ppi_2mode} displays the induced subgraph of two communities that \model detects. 2-mode communities detected by \model clearly reveal the interaction between different protein groups. For example, proteins in group $C$ of Figure~\ref{fig:ppi_2mode} heavily interact with proteins in group $A$, even though these proteins do not interact within the same group (with $A$ or within $C$).

To further analyze the role of these communities, we used gene ontologies to identify relevant terms/functions of proteins in $A$, $B$, $C$, and $D$ using the GO Term Finder~\cite{Boyle04GOTermFinder}. The proteins in the large groups ($C$, $D$) are generally associated with catalytic activity and ion binding ($p$-value $\sim 10^{-4}$).

However, these proteins are regulated by different protein groups ($A$, $B$) which have different functions. Proteins in $A$ (\eg, YLR347C and YNL189W) are protein transporters, whereas proteins in $B$ (\eg, YLR291C) are regulators. Perhaps more interestingly, YPL070W belongs to both $A$ and $B$ and regulates both $C$ and $D$. However, its role is not yet known. But based on known functions of proteins in groups $A$ and $B$ we can extrapolate the function of YPL070W. This example shows how network analysis and community detection in particular can provide research directions for experimental biology~\cite{Carney09VPS9}.


\begin{figure*}[tp]
\begin{center}
\includegraphics[width=0.99\textwidth]{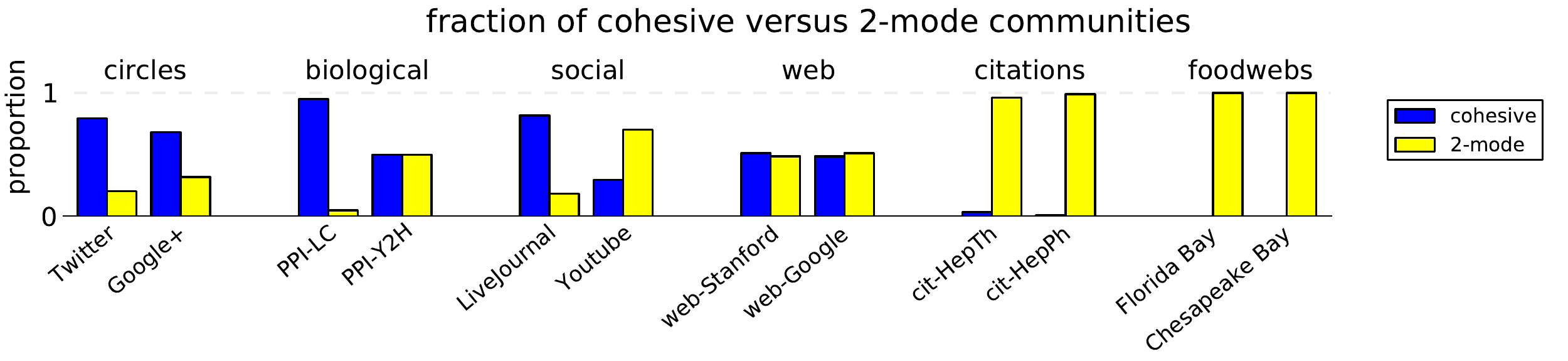}
\end{center}
\vspace{-8mm}
\caption{Fraction of 2-Mode communities and cohesive communities in six different types of networks. \label{fig:bar_chart}}
 \vspace{-2mm}
\end{figure*}

\subsection{2-mode vs. Cohesive Communities}
Since \model can detect both cohesive and 2-mode communities, we can use it to measure the extent to which real network data exhibits cohesive and 2-mode behavior.
This analysis allows us to characterize the mesoscale structure of real-world networks as the proportion of 2-mode versus cohesive communities can be used to gain further insights into community structure of networks.

\xhdr{Experimental setup}
For this experiment, we consider 12 networks from 6 domains in order to characterize their different community structures.
We consider ego networks (Twitter, Google+) and social networks (LiveJournal, Youtube) from Section~\ref{sec:experiments}.
We also include 8 networks from 4 different domains: Biological networks, web graphs, foodwebs, and citation networks among research papers from Table~\ref{tab:data-nogt}.


To classify whether a detected community is 2-mode or cohesive, we measure the Jaccard similarity $J(c) = \frac{|O(c) \cap I(c)|}{|O(c) \cup I(c)|}$ between the set of member nodes with outgoing memberships $O(c)$, and the set of member nodes with incoming memberships $I(c)$. In a completely cohesive community, this Jaccard similarity is 1 because two sets of members are identical, whereas it is 0 in a completely 2-mode community.
We regard a community $c$ as 2-mode if $J(c)$ is lower than some threshold $\gamma$ or as cohesive otherwise ($J(c) \geq \gamma$).
We use $\gamma = 0.2$ as this setting gives the most interpretable results.


\xhdr{Experimental results}
Figure~\ref{fig:bar_chart} shows the fraction of 2-mode and cohesive communities in 12 networks described above.
Ego networks (Twitter and Google+) exhibit a relatively high fraction of cohesive communities and as noted earlier Facebook ego networks (not shown) have an even higher fraction (over 95\%) of cohesive communities.
This result is in line with~\cite{demon} where the authors show that Facebook ego networks can be easily divided into cohesive communities. However, it is important to note that a significant fraction of Twitter (20\%) and Google+ (30\%) communities exhibit 2-mode structure. 


Literature-curated protein-protein interaction networks (PPI-LC) practically have only cohesive communities (and no 2-mode). On the other hand, in PPI networks generated based on yeast two-hybrid screening (PPI-Y2H) about 50\% of the communities are 2-mode.
This difference is interesting and confirms a previous study of PPI networks~\cite{Yu08YeastPPI}, which provided the following explanation:
Edges of PPI-LC are extracted from scientific papers that report experimentally validated interactions. However, current biological experiments have mainly been guided by research on cohesive communities and thus it seems as though most interactions that have been explored take place in ``cohesive'' communities~\cite{Yu08YeastPPI}. On the other hand, the PPI-Y2H network is created by a noisy automatic process and more faithfully represents the interaction network. In this case many 2-mode communities emerge~\cite{Yu08YeastPPI}.


In social networks we also find interesting results. In LiveJournal, communities are more cohesive, which can be explained by the fact that edges in LiveJournal indicate ``friendships'' (\ie, sharing private blog content).
On the other hand, Youtube communities are predominantly 2-mode. Youtube differs from other social networks in one important way: Edges in Youtube are essentially ``subscriptions'' for content rather than mutual friendships; consequently, high degree nodes tend to connect to low degree nodes~\cite{mislove07measurement}. 

Web graphs are of interest because Kumar et al.~\cite{KRRT99_trawling} used the existence of 2-mode communities as indicators or signatures for cohesive communities. Our results nicely suggest the co-existence of cohesive communities and 2-mode communities by showing that web graphs have an equal proportion of 2-mode and cohesive communities.

Finally, foodwebs as well as citation networks consist almost entirely of 2-mode communities.
These results are natural as reciprocal and cohesive relationships are extremely unlikely in these networks. In foodwebs, for example, few species prey upon each other. Citation networks are directed acyclic graphs and reciprocal citation is impossible by definition. Intuitively, cohesive communities in directed networks contain some number of bidirectional edges among their members, therefore a lack of such reciprocal edges naturally leads to the dominance of 2-mode communities, as we observe in Figure~\ref{fig:foodweb_2mode}.

\section{Conclusion}
\label{sec:conclusion}

An accurate notion of a \emph{community} is critical when studying the mesoscale structure of networks. Traditional models consider `communities' to be sets of densely connected nodes. In addition, here we also consider \emph{2-mode} communities, which are groups of nodes who may not link to each other but link in a coordinate way to the other nodes in the network.


We have presented \model, a community detection method which naturally detects both densely connected and 2-mode communities. \model can capture overlapping and hierarchical structure among communities, and handles both directed and undirected networks. Our experimental findings reveal that \model outperforms the current state-of-the-art in detecting ground-truth communities. Moreover, \model also reveals how 2-mode and cohesive communities co-exist in real networks.

The versatility of \model to detect both cohesive and 2-mode communities accurately in directed and undirected networks raises many interesting avenues of future work. For example, understanding the interaction between 2-mode communities and cohesive communities is a fruitful direction. Inferring the role of nodes from their community affiliations would also be useful.
%
Another idea is to extend \model to find important nodes in each community. This could be achieved by the fact that \model estimates real-valued membership strengths ($F_{uc}$ and $H_{uc}$) of each node to each community. From the values of $F_{uc}$ and $H_{uc}$ for node $u$ and community $c$, we could determine which nodes are most important and have the ``heaviest'' membership to a given community $c$.


\vspace{2mm}
\xhdr{Acknowledgements}
This research has been supported in part by NSF
IIS-1016909,              
CNS-1010921,              
CAREER IIS-1149837,       
IIS-1159679,              
ARO MURI,                 
DARPA GRAPHS,             
ARL AHPCRC,
Okawa Foundation,          
PayPal,
Docomo,                    
Boeing,                    
Allyes,                    
Volkswagen,                
Intel,                     
Alfred P. Sloan Fellowship, and    
the Microsoft Faculty Fellowship. 


\footnotesize
\begin{thebibliography}{10}
\vspace{2mm}

\bibitem{adamicBlogs}
L.~Adamic and N.~Glance.
\newblock The political blogosphere and the 2004 {U.S.} election: divided they
  blog.
\newblock In {\em LinkKDD '05}, 2005.

\bibitem{Ahn10LinkCommunitiesNature}
Y.-Y. Ahn, J.~Bagrow, and S.~Lehmann.
\newblock {Link communities reveal multi-scale complexity in networks}.
\newblock {\em Nature}, 2010.

\bibitem{airoldi07blockmodel}
E.~Airoldi, D.~Blei, S.~Fienberg, and E.~Xing.
\newblock Mixed membership stochastic blockmodels.
\newblock {\em JMLR}, 2007.

\bibitem{andersen06seed}
R.~Andersen and K.~Lang.
\newblock Communities from seed sets.
\newblock In {\em WWW '06}, 2006.

\bibitem{blockLDA}
R.~Balasubramanyan and W.~Cohen.
\newblock Block-{LDA}: Jointly modeling entity-annotated text and entity-entity
  links.
\newblock In {\em SDM '11}, 2011.

\bibitem{Ball11overlappingcommunities}
B.~Ball, B.~Karrer, and M.~Newman.
\newblock Efficient and principled method for detecting communities in networks.
\newblock In {\em Phys. Rev. E}, 2011.

\bibitem{Boyle04GOTermFinder}
E.~Boyle et al.
\newblock {GO::TermFinder}---open source software for accessing gene ontology
  information and finding significantly enriched gene ontology terms associated
  with a list of genes.
\newblock {\em {Bioinformatics}}, 2004.

\bibitem{breiger74groups}
L.~Breiger.
\newblock The duality of persons and groups.
\newblock {\em Social Forces}, 1974.

\bibitem{burt78}
R.~Burt.
\newblock Cohesion versus structural equivalence as a basis for network
  subgroups.
\newblock {\em Sociological Methods and Research}, 1978.

\bibitem{Carney09VPS9}
D.~Carney, B.~Davies, and B.~Horazdovsky.
\newblock {Vps9 domain-containing proteins: activators of Rab5 GTPases from
  yeast to neurons}.
\newblock {\em Trends in Cell Biology}, 2006.

\bibitem{demon}
M.~Coscia, G.~Rossetti, F.~Giannotti, and D.~Pedreschi.
\newblock Demon: a local-first discovery method for overlapping communities.
\newblock In {\em KDD '12}, 2012.

\bibitem{danon05community}
L.~Danon, J.~Duch, A.~Diaz-Guilera, and A.~Arenas.
\newblock Comparing community structure identification.
\newblock {\em J. of Stat. Mech.: Theory and Experiment}, 2005.

\bibitem{dhillon07graclus}
I.~Dhillon, Y.~Guan, and B.~Kulis.
\newblock Weighted graph cuts without eigenvectors: A multilevel approach.
\newblock {\em IEEE PAMI}, 2007.

\bibitem{fortunato09community}
S.~Fortunato.
\newblock Community detection in graphs.
\newblock {\em Physics Reports}, 2010.

\bibitem{Gleich12neighborhoods}
D.~Gleich and Seshadhri.
\newblock Neighborhoods are good communities.
\newblock In {\em KDD '12}, 2012.

\bibitem{mmsbScale12}
P.~Gopalan, D.~Mimno, S.~Gerrish, M.~Freedman, and D.~Blei.
\newblock Scalable inference of overlapping communities.
\newblock In {\em NIPS '12}, 2012.

\bibitem{Henderson12RolX}
K.~Henderson, B.~Gallagher, T.~Eliassi-Rad, H.~Tong, S.~Basu, L.~Akoglu,
  D.~Koutra, C.~Faloutsos, and L.~Li.
\newblock Rolx: structural role extraction {\&} mining in large graphs.
\newblock In {\em KDD}, 2012.

\bibitem{Holland83BlockModel}
P.~W. Holland, K.~B. Laskey, and S.~Leinhardt.
\newblock Stochastic blockmodels: First steps.
\newblock {\em Social Networks}, 1983.

\bibitem{Hsieh11FastNMF}
C.-J. Hsieh and I.~S. Dhillon.
\newblock Fast coordinate descent methods with variable selection for
  non-negative matrix factorization.
\newblock In {\em KDD '11}, 2011.

\bibitem{hierarchical}
S.~Johnson.
\newblock Hierarchical clustering schemes.
\newblock {\em Psychometrika}, 1967.

\bibitem{karypis98metis}
G.~Karypis and V.~Kumar.
\newblock Multilevel k-way partitioning scheme for irregular graphs.
\newblock {\em J. of Parallel and Distributed Computing}, 1998.

\bibitem{klimt04enron}
B.~Klimt and Y.~Yang.
\newblock Introducing the enron corpus.
\newblock In {\em CEAS '04}, 2004.

\bibitem{KRRT99_trawling}
R.~Kumar, P.~Raghavan, S.~Rajagopalan, and A.~Tomkins.
\newblock Trawling the web for emerging cyber-communities.
\newblock {\em Computer Networks}, 1999.

\bibitem{Kwak10Twitter}
H.~Kwak, C.~Lee, H.~Park, and S.~Moon.
\newblock What is twitter, a social network or a news media?
\newblock In {\em WWW '10}, 2010.

\bibitem{Lattanzi09AffiliationNetworks}
S.~Lattanzi and D.~Sivakumar.
\newblock Affiliation networks.
\newblock In {\em STOC '09}, 2009.

\bibitem{Newman08DirectedCommunity}
E.~Leicht and M.~E. Newman.
\newblock Community structure in directed networks.
\newblock {\em Phys. Rev. Lett.}, 2008.

\bibitem{jure05dpl}
J.~Leskovec, J.~Kleinberg, and C.~Faloutsos.
\newblock Graphs over time: densification laws, shrinking diameters and
  possible explanations.
\newblock In {\em KDD '05}, 2005.

\bibitem{jure08ncp2}
J.~Leskovec, K.~J. Lang, A.~Dasgupta, and M.~W. Mahoney.
\newblock Community structure in large networks: Natural cluster sizes and the
  absence of large well-defined clusters.
\newblock {\em Internet Mathematics}, 2009.

\bibitem{Lin07gradientforNMF}
C-J.~Lin.
\newblock Projected Gradient Methods for Nonnegative Matrix Factorization.
\newblock {\em Neural Computation}, 2007.

\bibitem{Malliaros13Survey}
F. Malliaros and M. Vazirgiannis.
\newblock Clustering and Community Detection in Directed Networks: A Survey.
\newblock {\em Physics Reports}, 2013.


\bibitem{nips2012}
J.~McAuley and J.~Leskovec.
\newblock Learning to discover social circles in ego networks.
\newblock In {\em NIPS '12}, 2012.

\bibitem{emailRoles}
A.~McCallum, X.~Wang, and A.~Corrada-Emmanuel.
\newblock Topic and role discovery in social networks with experiments on enron
  and academic email.
\newblock {\em JAIR}, 2007.

\bibitem{mislove07measurement}
A.~Mislove, M.~Marcon, K.~P. Gummadi, P.~Druschel, and B.~Bhattacharjee.
\newblock Measurement and analysis of online social networks.
\newblock In {\em IMC '07}, 2007.

\bibitem{palla05_OveralpNature}
G.~Palla, I.~Der\'{e}nyi, I.~Farkas, and T.~Vicsek.
\newblock Uncovering the overlapping community structure of complex networks in
  nature and society.
\newblock {\em Nature}, 2005.

\bibitem{papadopoulos11community}
S.~Papadopoulos, Y.~Kompatsiaris, A.~Vakali, and P.~Spyridonos.
\newblock Community detection in social media.
\newblock {\em DMKD}, 2011.


\bibitem{Pinkert10ProteinModules}
S.~Pinkert, J.~Schultz, and J.~Reichardt.
\newblock Protein interaction networks---more than mere modules.
\newblock {\em PLoS CompBio}, 2010.

\bibitem{RCCLP04_PNAS}
F.~Radicchi, C.~Castellano, F.~Cecconi, V.~Loreto, and D.~Parisi.
\newblock Defining and identifying communities in networks.
\newblock {\em PNAS}, 2004.

\bibitem{rosvall07_informationPNAS}
M.~Rosvall and C.~Bergstrom.
\newblock An information-theoretic framework for resolving community structure
  in complex networks.
\newblock {\em PNAS}, 2007.

\bibitem{rosvall08infomap}
M.~Rosvall and C.~T. Bergstrom.
\newblock Maps of random walks on complex networks reveal community structure.
\newblock {\em PNAS}, 2008.

\bibitem{Satuluri08Markov}
V. Satuluri and S. Parthasarathy.
\newblock Scalable Graph Clustering using Stochastic Flows: Applications to Community Discovery.
\newblock {\em KDD '09}, 2009.

\bibitem{ulanowicz98foodweb_baywet}
R.~Ulanowicz, C.~Bondavalli, and M.~Egnotovich.
\newblock Network analysis of trophic dynamics in south florida ecosystem, fy
  97: The florida bay ecosystem.
\newblock {\em Ref. CBL98-123. Chesapeake Biological Laboratory}, 1998.

\bibitem{Xie13SurveyOverlapping}
J.~Xie, S.~Kelley, and B.~K. Szymanski.
\newblock Overlapping community detection in networks: the state of the art and
  comparative study.
\newblock {\em ACM Computing Surveys}, 2013.

\bibitem{jaewon12agmfit}
J.~Yang and J.~Leskovec.
\newblock Community-affiliation network model for overlapping community
  detection.
\newblock In {\em ICDM '12}, 2012.

\bibitem{jaewon11comscore}
J.~Yang and J.~Leskovec.
\newblock Defining and evaluating network communities based on ground-truth.
\newblock In {\em ICDM '12}, 2012.

\bibitem{jaewon13agmfast}
J.~Yang and J.~Leskovec.
\newblock Overlapping community detection at scale: A non-negative
  factorization approach.
\newblock In {\em WSDM '13}, 2013.

\bibitem{Yu08YeastPPI}
H.~Yu, P.~Braun, M.~A. Yildirim, I.~Lemmens, K.~Venkatesan, J.~Sahalie,
  T.~Hirozane-Kishikawa, F.~Gebreab, N.~Li, N.~Simonis, and et~al.
\newblock High-quality binary protein interaction map of the yeast interactome
  network.
\newblock {\em Science}, 2008.

\bibitem{zheleva09affiliation}
E.~Zheleva, H.~Sharara, and L.~Getoor.
\newblock Co-evolution of social and affiliation networks.
\newblock In {\em KDD '09}, 2009.

\end{thebibliography}


\end{document}